\documentclass{IEEEtaes}

\usepackage{color,array,amsthm}
\usepackage{graphicx}
\usepackage{amsmath} 
\usepackage{mathrsfs}
\usepackage{amssymb}  
\usepackage{multirow}

\usepackage[labelformat=simple]{subcaption}
\jvol{XX}
\jnum{XX}
\jmonth{XXXXX}
\paper{1234567}
\pubyear{2024}
\doiinfo{TAES.2024.Doi Number}

\begin{document}

\clearpage
\onecolumn
\thispagestyle{empty}
\noindent
This paper has been accepted for publication in \textit{IEEE Transactions on Aerospace and Electronic Systems}.

\begin{center}
    DOI: \href{https://doi.org/10.1109/TAES.2025.3594088}{10.1109/TAES.2025.3594088}
    
    IEEE Xplore: \href{https://ieeexplore.ieee.org/document/11105436}{https://ieeexplore.ieee.org/document/11105436}
\end{center}
\vspace{2em}

\noindent
© 2025 IEEE.  Personal use of this material is permitted.  Permission from IEEE must be obtained for all other uses, in any current or future media, including reprinting/republishing this material for advertising or promotional purposes, creating new collective works, for resale or redistribution to servers or lists, or reuse of any copyrighted component of this work in other works.

\clearpage
\twocolumn
\setcounter{page}{1}

\title{Multi-frame Detection via Graph Neural Networks: A Link Prediction Approach}

\author{Zhihao Lin}
\member{Student Member, IEEE}
\affil{National Key Laboratory of Radar Signal Processing, Xidian University, Xi’an, 710071, China} 

\author{Chang Gao}
\member{Member, IEEE}
\affil{National Key Laboratory of Radar Signal Processing, Xidian University, Xi’an, 710071, China} 

\author{Junkun Yan}
\member{Senior Member, IEEE}
\affil{National Key Laboratory of Radar Signal Processing, Xidian University, Xi’an, 710071, China}

\author{Qingfu Zhang}
\affil{Department of Computer Science, City University of Hong Kong, Hong Kong}
\member{Fellow, IEEE}

\author{Bo Chen}
\member{Senior Member, IEEE}
\affil{National Key Laboratory of Radar Signal Processing, Xidian University, Xi’an, 710071, China}

\author{Hongwei Liu}
\affil{National Key Laboratory of Radar Signal Processing, Xidian University, Xi’an, 710071, China}
\member{Senior Member, IEEE}

\receiveddate{Manuscript received XXXXX 00, 0000; revised XXXXX 00, 0000; accepted XXXXX 00, 0000.\\
This work was supported in part by Hong Kong Innovation and Technology Commission Funding Administrative System II (ITF Ref. No. GHP/110/20GD) and the National Natural Science Foundation of China (62192714, U21B2006).{\itshape(Corresponding author: Hongwei Liu; Chang Gao.)}}

\authoraddress{
Authors’ addresses: Zhihao Lin, Junkun Yan, Bo Chen and Hongwei Liu are with the National Key Laboratory of Radar Signal Processing, Xidian University, Xi'an, 710071, China, E-mail:
(\href{mailto:zhihlin@stu.xidian.edu.cn}
{zhihlin@stu.xidian.edu.cn}; 
\href{mailto:jkyan@xidian.edu.cn}
{jkyan@xidian.edu.cn}; 
\href{mailto:bchen@mail.xidian.edu.cn}
{bchen@mail.xidian.edu.cn}; 
\href{mailto:hwliu@xidian.edu.cn}
{hwliu@xidian.edu.cn}). 
Chang Gao is with the National Key Laboratory of Radar Signal Processing, Xidian University, Xi'an, 710071, China, and was also with the Department of Computer Science, City University of Hong Kong, Hong Kong, E-mail: (\href{mailto:changgao@xidian.edu.cn}
{changgao@xidian.edu.cn}).
Qingfu Zhang is with the Department of Computer Science, City University of Hong Kong, Hong Kong, and also with the City University of Hong Kong Shenzhen Research Institute, Shenzhen, 518057, China,  E-mail: (\href{mailto:qingfu.zhang@cityu.edu.hk}
{qingfu.zhang@cityu.edu.hk}).
{\itshape(Corresponding author: Hongwei Liu; Chang Gao.)}
}

\markboth{Zhihao Lin ET AL.}{GLP-MFD}
\maketitle

\begin{abstract}
Multi-frame detection algorithms can effectively utilize the correlation between consecutive echoes to improve the detection performance of weak targets. 
Existing efficient multi-frame detection algorithms are typically based on three sequential steps: plot extraction via a relatively low primary threshold, track search, and track detection.
However, these three-stage processing algorithms may result in a notable loss of detection performance and do not fully leverage the available echo information across frames.
For the application of graph neural networks in multi-frame detection, the algorithms are primarily based on node classification tasks, which do not focus on directly outputting target tracks.
In this paper, we reformulate the multi-frame detection problem as a link prediction task in graphs. 
First, we perform a rough association of multi-frame observations that exceed the primary threshold to construct observation association graphs.
Subsequently, the multi-feature link prediction network is designed based on graph neural networks, which integrates multi-dimensional information, such as Doppler, signal structures, and spatio-temporal coupling of observations.
By leveraging the principle of link prediction, we unify the process of track search and track detection into one step to reduce performance loss and directly output target tracks.
Numerical results indicate that the proposed algorithm improves the detection performance of weak targets while suppressing false alarms, compared with traditional single-frame and multi-frame detection algorithms. 
Additionally, interpretability analysis shows that the designed network effectively integrates the utilized features, allowing for accurate target associations.

\end{abstract}


\section{INTRODUCTION}
T{\scshape arget} detection aims to determine whether a signal received by sensors originates from a target based on the signal strength \cite{ref1}.
Effective target detection serves as the foundation for subsequent operations such as target tracking and target recognition.
The reliable detection of weak targets still remains a critical challenge that urgently needs to be addressed in systems such as radar and sonar.
The traditional signal processing framework, which follows a sequential pattern of track after detection, struggles to balance detection performance with system processing load \cite{ref2},\cite{ref3},\cite{ref4}. 
Its drawbacks are particularly evident in weak target detection scenarios.
To address the shortcomings of traditional signal processing, multi-frame detection (MFD) methods leverage the regularity of target motion and the accumulation of target energy to improve the detection performance for low signal-to-noise ratio (SNR) targets \cite{ref5},\cite{ref7},\cite{ref_Orlando_1},\cite{ref_Orlando_2}.

Traditional MFD methods enable the successful detection of weak targets by directly utilizing un-thresholded raw data \cite{ref8},\cite{ref10},\cite{ref11},\cite{ref12},\cite{ref13}.
But they exhibit drawbacks in high-resolution detection scenarios that require a large number of detection cells.
On one hand, the large volume of observations will impose a substantial computational burden on the system.
On the other hand, the observation uncertainty introduced by excessive false alarms makes it difficult to perform accurate target associations.
Although some methods focus on improving the accuracy of target measurements in different scenarios, they still do not consider pre-filtering the candidate detection set based on the strength of the received echoes \cite{ref15},\cite{ref16},\cite{ref17}.
A viable approach to address the aforementioned challenge is a type of three-stage MFD methods, which typically involves three steps \cite{ref18},\cite{ref19},\cite{ref20}:
First, a relatively low primary threshold is set for detection and plot extraction on each frame. 
The low threshold aims to maintain good detection performance for weak targets while preliminarily filtering out observations unlikely to originate from targets, thereby reducing computational load. 
Second, for multi-frame observations that exceed the primary threshold, potential tracks are searched by leveraging target kinematic constraints. 
Finally, a fusion detection is made along the potential tracks based on information such as echo energy. 
This kind of methods transforms single-frame plot-level detection into multi-frame track-level detection, improving detection performance for weak targets.
However, such sequential track search and track detection hinder the overall performance by the limitations of the potential track search.

Traditional methods of track search primarily rely on position and Doppler information to establish kinematic constraints \cite{ref19},\cite{ref21},\cite{ref22},\cite{ref23}. 
In MFD methods that incorporate neural networks, reference \cite{ref24} adopts the three-stage scheme, first performing model-driven track search based on the position and Doppler measurements, then conducting track detection using a data-driven method \cite{ref25}.
Reference \cite{ref26} extends the scenario to the bistatic radar system based on this concept.
Nevertheless, the model-driven track search used in these methods often leads to target missed associations when there is a model mismatch. 
Because it is difficult to apply unified constraint parameters to reliably associate targets with different kinematic features, which in turn hinders the track detection process.

Graphs, as a mathematical language in the physical world, can reasonably describe multi-frame data \cite{ref27}. 
Compared to other neural networks, graph neural networks (GNNs) are capable of modeling data characteristics from both structural and functional perspectives, successfully extracting data features even with a limited number of labels \cite{ref28},\cite{ref29},\cite{ref30}.
Although the sparse distribution of targets in the observation space results in a limited number of target labels, the use of GNNs can effectively address this issue.
In the detection methods which apply GNNs, reference \cite{ref32} introduces a novel approach using GNNs to develop a message-passing solution for the inference task of massive multiple-input multiple-output (MIMO) detection in wireless communication.
Reference \cite{ref33} constructs graphs from point cloud data obtained from sensors in the field of autonomous driving.
The method leverages GNNs to fully exploit point features and relationship characteristics among points, achieving object detection and semantic segmentation through a node classification task. 
Reference \cite{ref34}, in a maritime scenario, similarly employs the node classification concept within GNNs to distinguish targets and clutter in multi-frame observations.
However, the utilization of the node classification task does not focus on directly outputting target tracks. 
Additionally, edge features can be considered for design by incorporating domain knowledge.

To overcome the limitations of model-driven track search and achieve target detection in environments with dense false alarms, a data-driven approach should enhance the accuracy of target associations \cite{ref35}. 
The process of track detection from multi-frame observations can be viewed as a nonlinear mapping from multi-dimensional information to the determination of whether a trajectory originates from a target. 
Furthermore, the track search and detection process is fundamentally equivalent to a clustering problem of spatio-temporal plot associations. 
Unlike the node classification, link prediction within graphs can be utilized to integrate multi-dimensional information on different attributes to accomplish track search and track detection in a unified manner \cite{ref36},\cite{ref37}. 
This approach allows for direct output of target trajectories from multi-frame observations, simplifying the processing logic while reducing the loss in track search, ultimately leading to improved target detection performance.

Based on the above considerations, we propose a multi-frame detection algorithm via graph neural network-based link prediction (GLP-MFD).
The proposed algorithm first performs a rough association for multi-frame observations that exceed the primary threshold.
And the successfully associated observations are constructed as observation association graphs. 
The features like Doppler information and signal structures of observations are represented in nodes, while the spatio-temporal coupling relationships among multi-frame observations are represented in edges. 
Then the multi-feature link prediction network (MFLPN) is designed to carry out the multi-frame detection process within observation association graphs. Our contributions are summarized as follows.

\begin{enumerate}
\def\labelenumi{\arabic{enumi})}
\item By modeling the multi-frame detection as a link prediction task in graphs, we overcome the performance bottleneck in traditional MFD methods, which arises from model mismatches and insufficient utilization of echo information during sequential track search and track detection. 
By constructing multi-frame observations as observation association graphs, we perform a link prediction task to directly output target tracks.
The proposed algorithm enables integrated track search and track detection, achieving superior target detection performance while ensuring effective suppression of false alarms.
\item By integrating information such as Doppler, signal structures, and spatio-temporal coupling, we specifically design MFLPN which effectively exploits multi-dimensional features.
Additionally, we leverage a unified cost function to enable the network to perform the link prediction task with higher accuracy.
\item We design edge features based on coupled position and Doppler measurements. 
A novel message passing function is designed to integrate edge features while incorporating an attention mechanism. 
This approach effectively extracts spatio-temporal coupling information among observations, thereby enhancing the network performance.
\end{enumerate}

The rest of this article is organized as follows.
Section \ref{A} briefly describes the MFD problem.
Sections \ref{B} and \ref{C} introduce the design of GLP-MFD and its corresponding computational complexity analysis, respectively. 
Section \ref{D} presents numerical results to validate the performance improvement of the proposed algorithm. 
Finally, Section \ref{E} concludes this article.

\section{PROBLEM FORMULATION}\label{A}
Coherent systems can improve the target SNR through pulse compression and coherent integration, while obtaining Doppler measurements. 
After detection and plot extraction with the primary threshold $\gamma_1$, the $k$-th observation at frame $n$ can be represented as a vector:
\begin{equation}
\boldsymbol{z}_{k, n}=\left(t_{k, n}, r_{k, n}, \theta_{k, n}, v_{k, n}, \boldsymbol{a}_{k, n}\right)^{\top}\text{,}
\label{eq1}
\end{equation}
where $t_{k, n}$ is the time instant when the observation is taken, $r_{k, n}$ is the range measurement, $\theta_{k, n}$ is the azimuth measurement, $v_{k, n}$ is the potentially ambiguous radial velocity, $\boldsymbol{a}_{k, n}$ is the range-Doppler map and $\top$ denotes the transpose operation. 
For better readability, we consider a 2-D scenario.
The relationship between the true target state $\left(x_{k, n}, \dot{x}_{k, n}, y_{k, n}, \dot{y}_{k, n}\right)$ in the Cartesian coordinate and the sensor measurements is as follows:
\begin{equation}
\begin{aligned}
r_{k, n} & =\sqrt{x_{k, n}^2+y_{k, n}^2}+w_{k, n}^r\text{,} \\
\theta_{k, n} & =\arctan \frac{y_{k, n}}{x_{k, n}}+w_{k, n}^\theta\text{,} \\
v_{k, n} & =\left(\frac{\dot{x}_{k, n} \cdot x_{k, n}+\dot{y}_{k, n} \cdot y_{k, n}}{\sqrt{x_{k, n}^2+y_{k, n}^2}}+\frac{v_u}{2}\right) \bmod v_u \\
& -\frac{v_u}{2}+w_{k, n}^v\text{,}
\end{aligned}
\end{equation}
where $v_u$ denotes the range of the unambiguous radial velocity. $w_{k, n}^r$, $w_{k, n}^\theta$ and $w_{k, n}^v$ represent the noise of range, azimuth and radial velocity measurements, respectively.
All three are zero-mean additive Gaussian white noise, with standard deviations proportional to the range, azimuth and velocity resolution, and inversely proportional to the square root of SNR.
Considering range ambiguity can be resolved by using techniques such as frequency stepping or transmitting orthogonal waveforms, we only focus on the potential ambiguity in radial velocity. The observation list corresponding to $D_{n}$ plots obtained in frame $n$ is defined as follows:
\begin{equation}
\mathbf{Z}_n=\left(\boldsymbol{z}_{1, n}, \boldsymbol{z}_{2, n}, \ldots, \boldsymbol{z}_{D_n, n}\right)\text{.}
\end{equation}

The task of multi-frame target detection is to determine whether a target exists within the multi-frame observations $\left\{\mathbf{Z}_l\right\}_{l=n}^{n+L-1}$, where $L$ represents the number of frames processed jointly, also known as the length of the processing sliding window. 
In traditional three-stage MFD methods, a model-driven track search is first preformed based on target kinematic constraints:
\begin{equation}
\Lambda \left(\mathbf{Z}_n, \mathbf{Z}_{n+1},..., \mathbf{Z}_{n+L-1} \right) \leq \boldsymbol{\mathit{\Upsilon}} \text{,}
\end{equation}
where $\Lambda$ represents the function of constraints and $\boldsymbol{\mathit{\Upsilon}}$ represents the vector of constraint parameters.
The multi-frame observations meeting the constraints will be associated to form potential target tracks.
And those failing to meet the constraints will be regarded as false alarms.
Then, a detector for potential tracks can be designed using methods such as the generalized likelihood ratio test \cite{ref38},\cite{ref39},\cite{refdet2},\cite{refdet3}.
The potential tracks with test statistic exceeding the final detection threshold $\gamma_2$ will be declared as confirm tracks after pruning.
However, such model-driven track search usually suffers from a performance loss due to target missed associations.
And the detector is hard to derive when fusing multi-dimensional information on different attributes.
Therefore, a data-driven method is proposed in this work to address these problems.

\begin{figure*}
\centerline{\includegraphics[width=0.83\textwidth]{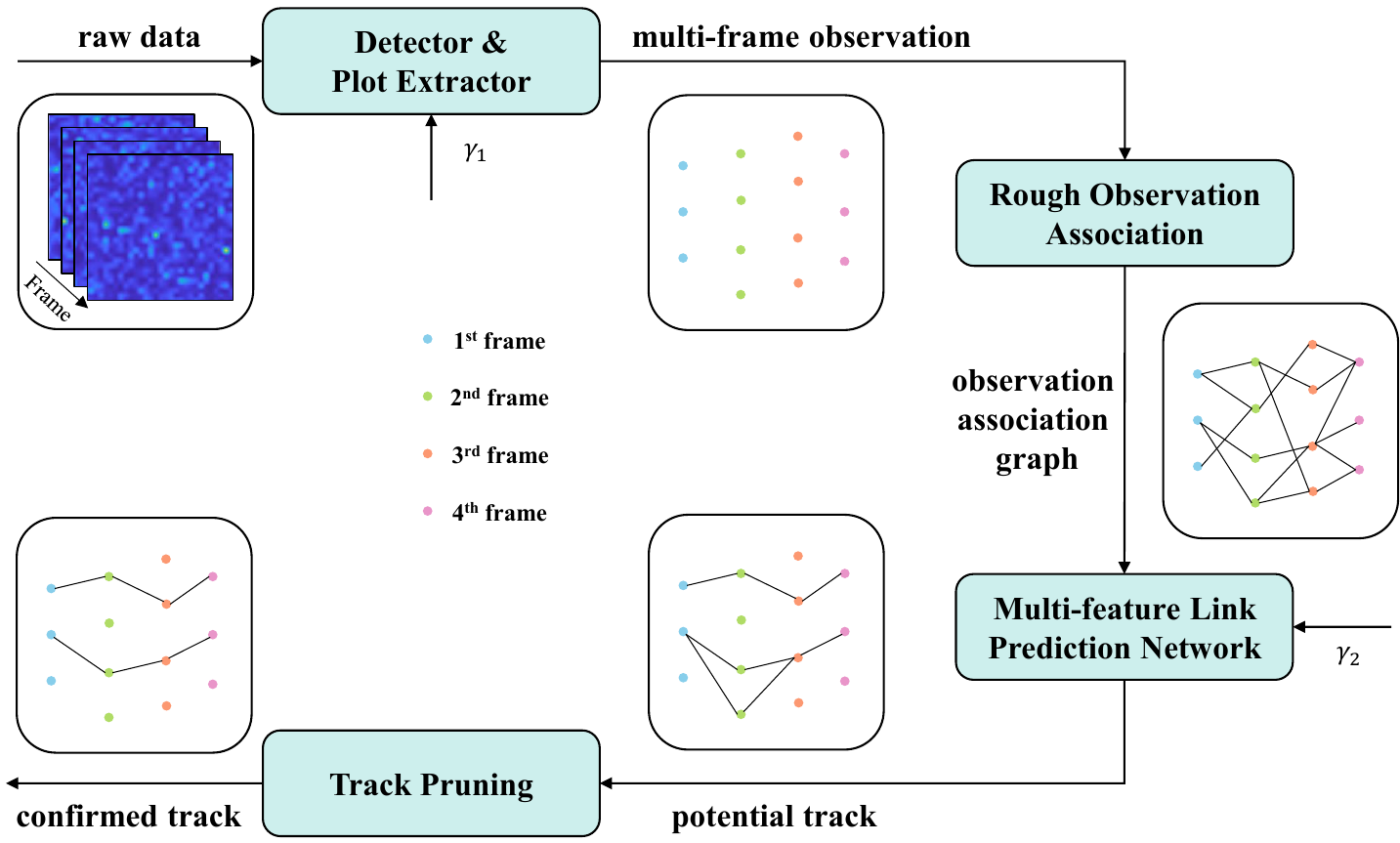}}
\caption{Operating scheme of GLP-MFD.}
\label{fig1}
\end{figure*}

\section{MULTI-FRAME DETECTION VIA GRAPH NEURAL NETWORK-BASED LINK PREDICTION}\label{B}
The operating scheme of the proposed GLP-MFD algorithm is depicted in Fig. \ref{fig1}.
First, detection and plot extraction are performed on the raw data using the primary threshold $\gamma_1$. 
To enhance the detection capability for weak targets, $\gamma_1$ is typically set to a relatively low value.
Additionally, it is essential to preliminarily eliminate false alarms to prevent imposing an undue computational burden on the system.
Subsequently, multi-frame observations are roughly associated by applying loose constraints to construct observation association graphs.
This step aims to enable target observations to readily meet the constraints, ensuring successful associations and preventing performance degradation due to association failure.
Then, the designed MFLPN is employed to complete the integrated track search and track detection process within the observation association graphs.
The potential tracks can be obtained after the network processing with the final detection threshold $\gamma_2$.
Finally, confirmed tracks are output after track pruning, which can be used for subsequent operations such as target tracking.

By employing the aforementioned data-driven approach, the designed network effectively learns the complex nonlinear mapping from multi-frame information to target detection.
In addition, the proposed algorithm integrates the two sequential steps of track search and track detection into one, thereby simplifying the processing logic while reducing loss. 
Furthermore, it effectively leverages multi-dimensional information on different attributes, leading to superior performance in detecting weak targets.

\subsection{Construction of observation association graphs}\label{subsec:3A}
An undirected graph $\mathcal{G}=(\mathcal{V}, \mathcal{E})$, referred to as the observation association graph is constructed by using a loose constraint. 
$\mathcal{V}$ denotes the set of nodes representing multi-frame observations and $\mathcal{E}$ signifies the set of edges that connect successfully associated nodes.
Correct target tracks exist within the graph and we aim to accurately find target tracks while eliminating false tracks. 
Due to the limited computational power of the system, it is unreasonable to associate all observations across frames. 
The velocities of the targets of interest typically have an upper bound.
We leverage the velocity limit to ensure successful target associations while roughly excluding false observations unlikely to be associated with targets.
In this way, we mitigate performance loss due to target missed associations and reduce the computational burden.
Therefore, observation association graphs are constructed by using the maximum velocity constraint.
For two observations $\left(t_1, r_1, \theta_1, v_1\right)$ and $\left(t_2, r_2, \theta_2, v_2\right)$ from different frames, where $t_1 < t_2$, the maximum velocity constraint is as follows:
\begin{equation}
\begin{aligned}
\| \mathbf{p}_2 &- \mathbf{p}_1 \| \leq v_{\max} (t_2 - t_1) \text{,} \\
\mathbf{p}_1 & = \left(r_1 \cos \theta_1, r_1 \sin \theta_1 \right) \text{,} \\
\mathbf{p}_2 & = \left(r_2 \cos \theta_2, r_2 \sin \theta_2 \right) \text{,}
\end{aligned}
\end{equation}
where $v_{\max }$ denotes the maximum velocity of the targets of interest and $\|\cdot\|$ denotes the magnitude.
The observations in $\left\{\boldsymbol{Z}_l\right\}_{l=n}^{n+L-1}$ that satisfy the maximum velocity constraint will be associated.

\begin{figure}
\centerline{\includegraphics[width=0.5\textwidth]{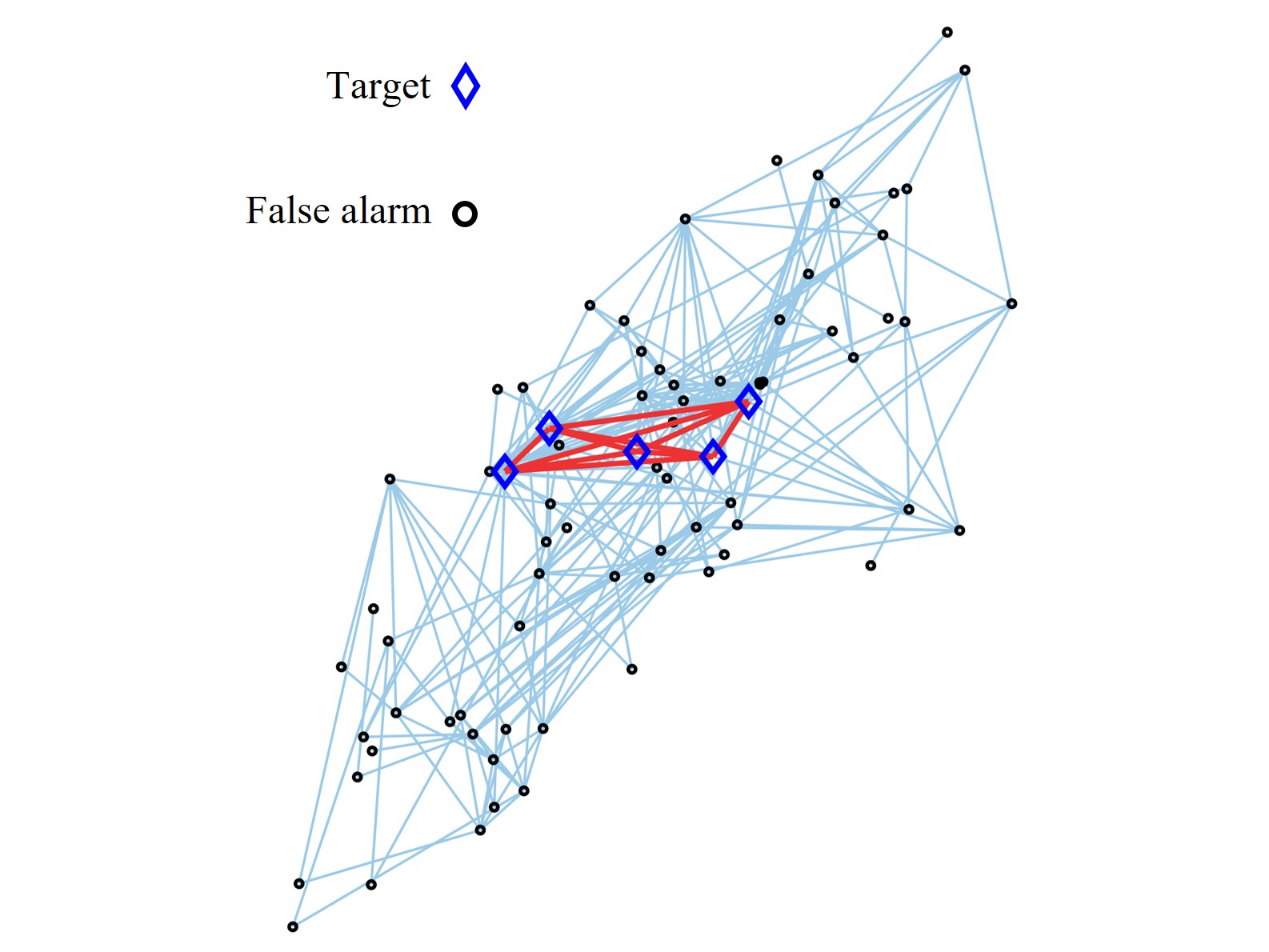}}
\caption{Schematic of the observation association graph.}
\label{fig2}
\end{figure}

Considering the difficulty in reliably detecting weak targets in each frame, as well as the need to ensure successful target associations and track continuity, we stipulate that observations need to undergo association testing with those in subsequent $Q$ frames.
Correspondingly, the maximum number of consecutive missed observations allowed for a track is set to $Q-1$.
If a track has $Q$ or more consecutive frames without observations, it is deemed to be of poor quality and should be discarded. 
An illustration of the constructed observation association graph is shown in Fig. \ref{fig2}.
The nodes represent multi-frame observations which exceed the primary threshold.
The edges represent observations undergoing association testing that satisfy the maximum velocity constraint.
The illustration demonstrates the successful associations among target observations, which helps to avoid performance loss due to missed associations. 
Additionally, the number of observations in the confirmed track is set to be at least $M$.
In other words, any path in the graph with $M$ or more nodes is considered as a candidate track.
And the link prediction acts on these candidate tracks to eliminate associations containing false alarms while preserving the associations between targets, thereby achieving integrated track search and track detection.

\subsection{Multi-feature link prediction network}
\begin{figure*}
\centerline{\includegraphics[width=0.83\textwidth]{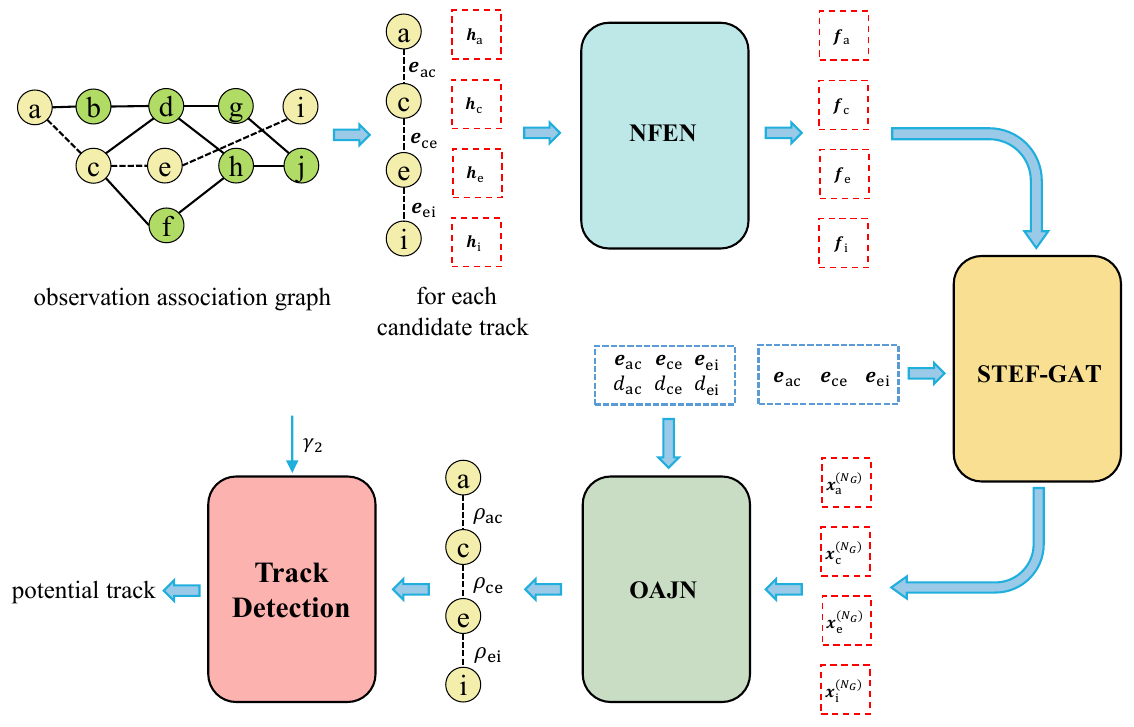}}
\caption{The architecture of the multi-feature link prediction network.}
\label{fig3}
\end{figure*}

Considering the challenges in deriving a detector that integrates multi-dimensional information, we address the problem from a data-driven perspective \cite{refdet4}.
By cleverly designing networks and the cost function, we perform fusion detection by comprehensively utilizing diverse information. 
The architecture of MFLPN is illustrated in Fig. \ref{fig3}.
After constructing observation association graphs, the node feature embedding network (NFEN) is developed to extract plot features such as Doppler information and signal structures by leveraging the strong fitting capabilities of deep neural networks (DNNs) for nonlinear mapping \cite{ref40}. 
Subsequently, we design the graph attention network with spatio-temporal coupling encoded in edge features (STEF-GAT) based on GNNs.
By incorporating an attention mechanism and constructing edge features during message passing, the network adequately extracts relational information that embodies physical motion patterns among target observations.
Then, we develop the observation association judgment network (OAJN) to conduct the link prediction task in graphs, yielding confidence for edges deriving from target associations.
And the confidence score for each candidate track originating from targets can be obtained.
Finally, based on the track confidence, the final detection threshold $\gamma_2$ is employed to make detection for candidate tracks, with those exceeding $\gamma_2$ declared as potential tracks.

\begin{figure}
\centerline{\includegraphics[width=0.45\textwidth]{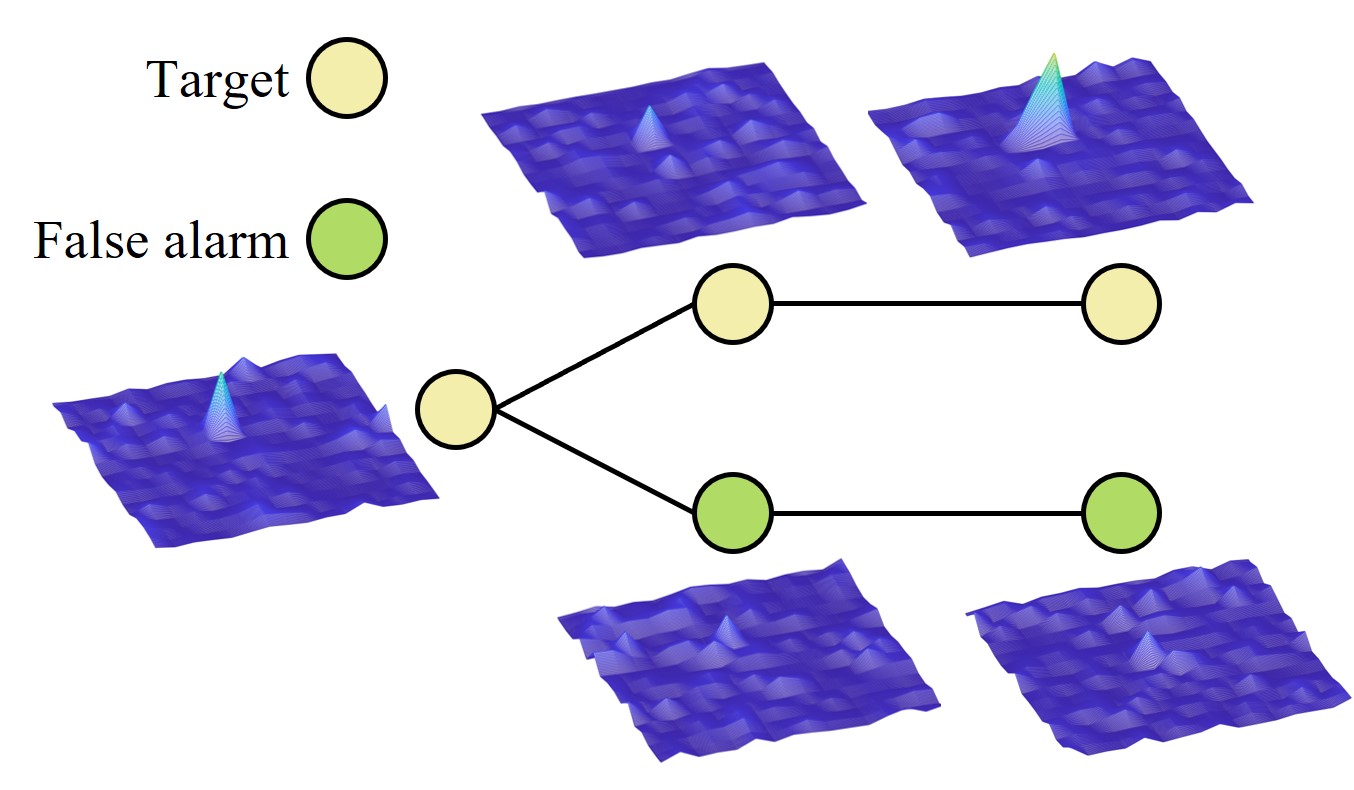}}
\caption{Illustration of the signal structure difference between targets and false alarms.}
\label{fig4}
\end{figure}

1) \textit{NFEN:}
Information such as the Doppler and signal structure can be obtained in a coherent system, while fully integrating the information of the target distributed in the echo should be helpful to detect the target from the background \cite{refdet1},\cite{ref41},\cite{ref42}.
Based on \eqref{eq1}, all the observation information we use is as follows:
\begin{equation}
\boldsymbol{z}_{k, n}=\left(t_{k, n}, r_{k, n}, \theta_{k, n}, v_{k, n}, d_{k, n}, s_{k, n}, \boldsymbol{i}_{k, n}, \boldsymbol{a}_{k, n}\right)^{\top}\text{,}
\end{equation}
where $d_{k, n}$ represents the Doppler channel in which the observation is detected, 
$s_{k, n}$ represents SNR, 
$\boldsymbol{i}_{k, n}$ represents temporal information, 
and $\boldsymbol{a}_{k, n}$ represents the range-Doppler map.
The variation of the Doppler channel $d_{k, n}$ where the target observation is located across multiple frames reflects the target kinematic characteristics.
For example, when a target moves at a constant speed, its Doppler channel remains stable or varies approximately linearly. 
$s_{k, n}$ not only reflects the strength of the target signal but also indicates the magnitude of measurement errors.
Therefore, the information contained in SNR is valuable and warrants thorough exploration.
The temporal information represents the chronological order of the observations in the track, serving as important temporal scale information. 
In this paper, the frame $n$ is converted to binary encoding as a temporal feature, denoted as $\boldsymbol{i}_{k, n}$.
As shown in Fig. \ref{fig4}, the signal structures of target observations across multiple frames exhibit a certain similarity, which is an important feature for distinguishing targets and false alarms. 
After coherent integration, the corresponding range-Doppler maps which contain signal structure information for specific detection cells can be obtained. 
We leverage range-Doppler maps to further extract characteristics of targets in the signal domain, thereby enhancing the separability between targets and false alarms.

The NFEN aims to perform node feature extraction on $\boldsymbol{h}_{k, n}=\left(d_{k, n}, s_{k, n}, \boldsymbol{i}_{k, n}, \boldsymbol{a}_{k, n}\right)^{\top}$, which enables the whole network to better fuse multi-dimensional information. 
It contains three fully connected neural networks and one convolutional neural network.
The fully connected layer is usually used for transformation of feature dimensions. 
The convolutional structure not only reduces the number of parameters through parameter sharing but also improves the learning ability of NFEN for local similarity information, which in turn leads to more effective extraction of plot features.

The transformations from $d_{k, n}, s_{k, n}, \boldsymbol{i}_{k, n}$ to their feature vectors $\boldsymbol{f}_{k, n}^{(d)}, \boldsymbol{f}_{k, n}^{(s)}, \boldsymbol{f}_{k, n}^{(\boldsymbol{i})}$ are achieved using fully connected neural networks:
\begin{equation}
\begin{aligned}
\boldsymbol{f}_{k,n}^{(d)}&=\left(D_d^{\left(N_d\right)} \circ \cdots \circ D_d^{(1)}\right)\left(d_{k, n}\right) \\
& =\phi_D\left(\mathbf{W}_d^{\left(N_d\right)} \cdots \phi_D\left(\mathbf{W}_d^{(1)} d_{k, n}\right)\right) \text{,} \\
\boldsymbol{f}_{k,n}^{(s)}&=\left(D_s^{\left(N_s\right)} \circ \cdots \circ D_s^{(1)}\right)\left(s_{k, n}\right) \\
& =\phi_D\left(\mathbf{W}_s^{\left(N_s\right)} \cdots \phi_D\left(\mathbf{W}_s^{(1)} s_{k, n}\right)\right) \text{,} \\
\boldsymbol{f}_{k,n}^{(\boldsymbol{i})}&=\left(D_{\boldsymbol{i}}^{\left(N_{\boldsymbol{i}}\right)} \circ \cdots \circ D_{\boldsymbol{i}}^{(1)}\right)\left(\boldsymbol{i}_{k, n}\right) \\
& =\phi_D\left(\mathbf{W}_{\boldsymbol{i}}^{\left(N_{\boldsymbol{i}}\right)} \cdots \phi_D\left(\mathbf{W}_{\boldsymbol{i}}^{(1)} \boldsymbol{i}_{k, n}\right)\right)\text{,}
\end{aligned}
\end{equation}
where $D_\cdot^\cdot$ denotes the transformations between layers,  $\circ$ denotes the composition of functions, $N_d, N_s, N_{\boldsymbol{i}}$ denote the number of fully connected layers, $\mathbf{W}_d^\cdot, \mathbf{W}_s^\cdot, \mathbf{W}_{\boldsymbol{i}}^\cdot$ denote the learnable parameter matrixes and $\phi_D$ denotes the nonlinear activation function in NFEN, which enhances the ability to learn nonlinear mappings.
The transformation from $\boldsymbol{a}_{k,n}$ to its feature vector $\boldsymbol{f}_{k,n}^{(\boldsymbol{a})}$ is completed using a convolutional neural network:
\begin{equation}
\begin{aligned}
\boldsymbol{f}_{k, n}^{(\boldsymbol{a})} & =\left(D_{\boldsymbol{a}}^{\left(N_{\boldsymbol{a}}\right)} \circ \cdots \circ D_{\boldsymbol{a}}^{(1)}\right)\left(\boldsymbol{a}_{k, n}\right) \\
& =\phi_D\left(\mathbf{W}_{\boldsymbol{a}}^{\left(N_{\boldsymbol{a}}\right)} * \cdots * \phi_D\left(\mathbf{W}_{\boldsymbol{a}}^{(1)} * \boldsymbol{a}_{k, n}\right)\right)\text{,}
\end{aligned}
\end{equation}
where $N_{\boldsymbol{a}}$ denotes the number of convolutional layers, $*$ denotes
the circular convolution operation and $\mathbf{W}_{\boldsymbol{a}}^\cdot$ denotes the parameters in the convolution kernels.
Finally, the concatenation of $\boldsymbol{f}_{k, n}^{(d)}, \boldsymbol{f}_{k, n}^{(s)}, \boldsymbol{f}_{k, n}^{(\boldsymbol{i})}, \boldsymbol{f}_{k, n}^{(\boldsymbol{a})}$ yields the multi-dimensional fused feature of the observation:
\begin{equation}
\boldsymbol{f}_{k, n}=\left[\boldsymbol{f}_{k, n}^{(d)} ; \boldsymbol{f}_{k, n}^{(s)} ; \boldsymbol{f}_{k, n}^{(\boldsymbol{i})}; \boldsymbol{f}_{k, n}^{(\boldsymbol{a})}\right]\text{,}
\end{equation}
where the fused feature $\boldsymbol{f}_{k, n}$ is used as the input node feature of the STEF-GAT in observation association graphs.
To enhance readability, we reformulate the input node features based on the temporal order of the observations across multiple frames. Specifically, if there are $D_l$ observations in the $l$-th frame across $L$ frames, $l=1,2, \ldots, L$, the input node features are rewritten as $\boldsymbol{x}_1^{(0)},\boldsymbol{x}_2^{(0)}, \ldots, \boldsymbol{x}_{\mathcal{L}}^{(0)}$, where $\mathcal{L}=\sum_{l=1}^L D_l$.

2) \textit{STEF-GAT:}
Most existing GNNs can be abstractly summarized into a universal framework known as message passing neural networks (MPNNs) \cite{ref43}. 
MPNNs focus on a specific message passing mechanism, allowing for the transfer of node information by defining a message passing function.
Therefore, MPNNs enable each node to aggregate information from its neighbors and update its state accordingly. 
The design of contemporary GNNs mostly adheres to this message passing concept.

\begin{figure*}
\centerline{\includegraphics[width=0.9\textwidth]{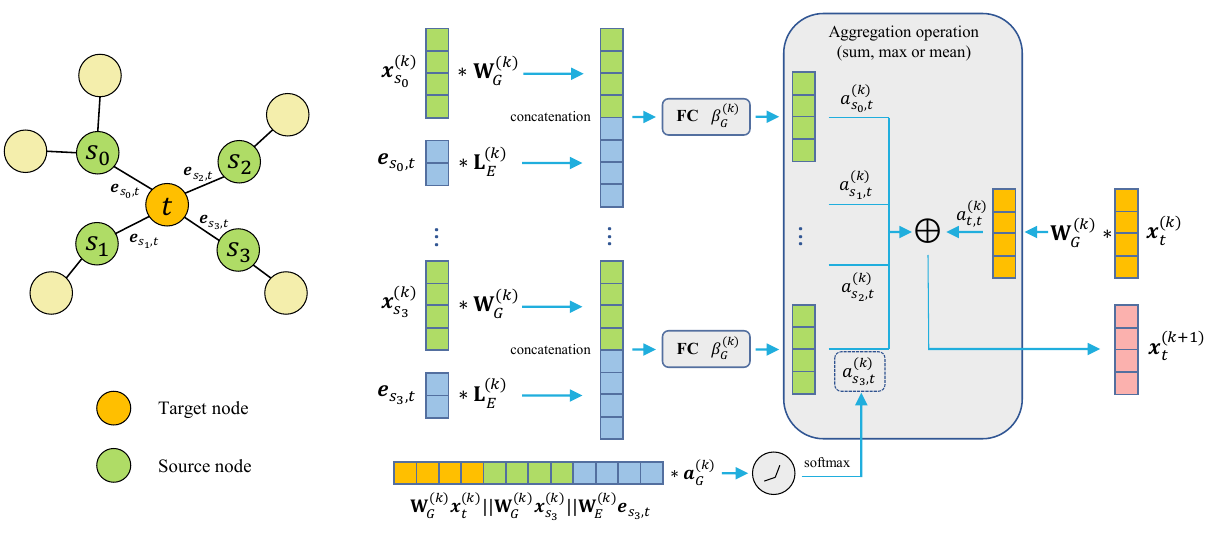}}
\caption{Message passing in STEF-GAT.}
\label{fig_add_1}
\end{figure*}

The observation association graphs constructed from data received by sensors at different times are distinct, indicating that the problem we address is an inductive task. 
Introducing an attention mechanism allows the network to focus more on relevant information that is crucial for decision-making, enabling efficient resource utilization and enhancing the generalization capability of the network, particularly performing well in inductive tasks \cite{ref44},\cite{ref45}. 
Furthermore, incorporating edge features facilitates a more thorough exploitation of the spatio-temporal coupling information among multi-frame observations, thereby improving the network performance. 
Based on these considerations, the message passing formula of the $k$-th layer in STEF-GAT is designed as follows:
\begin{equation}
\begin{aligned}
\boldsymbol{x}_i^{(k+1)} = & \hspace{0.1cm} \phi_G\left(a_{i, i}^{(k)}\mathbf{W}_G^{(k)} \boldsymbol{x}_i^{(k)} \underset{j \in \mathcal{N}(i)}{\oplus}\right.\\
& \hspace{-0.9cm}\left. 
\textcolor{white}{\underset{j \in \mathcal{N}(i)}{\oplus}}
a_{j, i}^{(k)} \beta_G^{(k)}\left(\mathbf{W}_G^{(k)} \boldsymbol{x}_j^{(k)} \| \hspace{0.05cm} \mathbf{L}_E^{(k)} \boldsymbol{e}_{j, i}\right)\right) \text{,}\\
a_{j, i}^{(k)} = & \hspace{0.1cm}
\frac{\exp \left(M_{j, i}^{(k)}\right)}{\sum_{m \in \mathcal{N}(i)\cup\{i\}} \exp \left(M_{m, i}^{(k)}\right)} \hspace{0.1cm} \text{,} \\
M_{m, i}^{(k)} = & \hspace{0.1cm}
\operatorname { L e a k y R e l u } \left(\boldsymbol{a}_G^{(k)} \left(\mathbf{W}_G^{(k)}\boldsymbol{x}_i^{(k)} \| \right.\right. \\
& \left.\left. \mathbf{W}_G^{(k)} \boldsymbol{x}_m^{(k)} \| \hspace{0.05cm} \mathbf{W}_E^{(k)} \boldsymbol{e}_{m, i}\right)\right)\text{.}
\end{aligned}
\label{eq10}
\end{equation}
$\boldsymbol{x}_i^{(k)}$ denotes the feature of node $i$ in the $k$-th layer.
And the input node feature matrix of STEF-GAT represented by $\boldsymbol{x}^{(0)}$ is the output fused feature matrix of NFEN represented by $\boldsymbol{f}$.
$\mathcal{N}(i)$ denotes the set of neighboring nodes of $i$ and $\boldsymbol{e}_{j, i}$ denotes the feature of the edge between node $i$ and node $j$.
$\mathbf{W}_G^{(k)}$ and $\mathbf{L}_E^{(k)}$ are the learnable matrixes of nodes and edges, respectively.
$\phi_G$ denotes the nonlinear activation function in STEF-GAT, such as ReLU, sigmoid and softmax.
$\beta_G^{(k)}$ represents a fully connected layer and $\|$ is a concatenation operation.
$\oplus$ represents an aggregation operation, such as summation, maximization, or averaging, which integrates the information from the neighboring nodes of $i$ into itself.
In the attention mechanism, $a_{j, i}^{(k)}$ represents the attention coefficient at the $k$-th layer, while $\boldsymbol{a}_G^{(k)}$ represents the learnable parameters and $\mathbf{W}_E^{(k)}$ represents the learnable matrix for edges. 
Unlike $\mathbf{L}_E^{(k)}$ which focuses on enhancing edge features, $\mathbf{W}_E^{(k)}$ emphasizes learning the impact of edge features within the attention mechanism. 
As shown in Fig. \ref{fig_add_1}, we illustrate the message passing in a STEF-GAT layer \cite{ref_add_1}. 
This demonstrates how information from source nodes, which are neighbors of the target node, is aggregated to the target node through message passing.
Additionally, employing multi-head attention mechanism can enhance the stability of the training process and improve the network accuracy \cite{ref46},\cite{ref47}. 
In this case, the message passing formula is rewritten as follows:
\begin{equation}
\boldsymbol{x}_i^{(k+1)}=\underset{h=1}{\overset{H}{\|}} \boldsymbol{x}_i^{(k+1, h)} \hspace{0.1cm} \text{.}
\end{equation}
$H$ represents the number of heads.
$\boldsymbol{x}_i^{(k+1, h)}$ represents the feature of node $i$ in the $h$-th head, which independently executes the transformation shown in \eqref{eq10}.
It is precisely that the message passing mechanism which enables STEF-GAT to integrate multi-frame spatial-temporal information, reveals the relationships among multi-frame observations, thereby enhancing the separability of targets and false alarms.

The time, range, azimuth and radial velocity measurements of target observations exhibit a certain coupling relationship across frames, reflecting the target physical kinematics. 
However, this coupling relationship is often complex.
Directly using $t_{k, n}, r_{k, n}, \theta_{k, n}, v_{k, n}$ as the node feature input to the network will require deeper network layers, increasing both the training difficulty and computational burden. 
To address this issue, we pre-construct the mapping relationships of observation measurements across multiple frames and treat the relationships as edge features. 
This method effectively extracts the spatio-temporal coupling information among multi-frame observations while reducing computational load and maintaining good performance.
In practice, target motion is typically based on a certain physical mechanism. 
For example, considering a target moving at a constant speed, it is possible to use the position measurements in two frames to estimate the target velocity. 
By projecting this velocity onto the sensor line-of-sight directions in each frame, the estimation of target radial velocities in both frames can be derived.
If there are no measurement errors, the estimated velocities should be equal to the velocity measurements. 
Considering the presence of errors, there will be a certain deviation between the estimations and the measurements, which can be used to determine whether two observations correspond to the same target.
Therefore, we incorporate the deviation as the edge feature in STEF-GAT to learn the confidence that two observations originate from the same target.

For two observations $\left(t_1, r_1, \theta_1, v_1\right)$ and $\left(t_2, r_2, \theta_2, v_2\right)$ from different frames, where $t_1 < t_2$, the observed radial velocities can be expressed as follows while taking into account the potential radial velocity ambiguities.
\begin{equation}
\begin{aligned}
v_1 + m_1 v_u, & \quad m_1 \in \mathbb{Z} \text{,}\\
v_2 + m_2 v_u, & \quad m_2 \in \mathbb{Z} \text{,}
\end{aligned}
\end{equation}
where $v_u$ denotes the range of the unambiguous radial velocity, while $m_1$ and $m_2$ denote the ambiguous numbers.
The target estimated radial velocities can be obtained based on the position measurements:
\begin{equation}
\begin{aligned}
& \hat{v}_1=\frac{r_2 \cos \left(\theta_2-\theta_1\right)-r_1}{t_2-t_1} \text{,} \\
& \hat{v}_2=\frac{r_2-r_1 \cos \left(\theta_2-\theta_1\right)}{t_2-t_1} \text{.}
\end{aligned}
\end{equation}
For the same target, there is a certain deviation between the estimated radial velocities and the observed radial velocities. 
We incorporate the deviation as a feature input to the network \cite{ref35}:
\begin{equation}
\begin{aligned}
F_{E, 1} &=\left|v_1+m_1 v_u-\hat{v}_1\right| \text{,}\\
F_{E, 2} &=\left|v_2+m_2 v_u-\hat{v}_2\right| \text{,}\\
m_1 &=\underset{m \in \mathbb{Z}}{\arg \min }\left|v_1+m_1 v_u-\hat{v}_1\right| \text{,}\\
m_2 &=\underset{m \in \mathbb{Z}}{\arg \min }\left|v_2+m_2 v_u-\hat{v}_2\right| \text{.}
\end{aligned}
\end{equation}
Since $ F_{E, 1}$ and $ F_{E, 2}$ are features derived from the information of two nodes, they can serve as edge features in observation association graphs, denoted as $\boldsymbol{e}_{1,2}=\left[F_{E, 1}, F_{E, 2}\right]$.

After obtaining the input features of nodes and edges for STEF-GAT, it leverages multi-layer message passing to extract information from multiple frames:
\begin{equation}
\boldsymbol{X}^{\left(N_G\right)}=\left(G^{\left(N_G\right)} \circ \ldots \circ G^{(2)} \circ G^{(1)}\right)(\boldsymbol{x}^{(0)}, \boldsymbol{e}) \hspace{0.1cm} \text{,}
\end{equation}
where $G^\cdot$ denotes the transformations between layers and $N_G$ is the number of layers. 
$\boldsymbol{x}^{(0)}$ represents the node feature matrix input to STEF-GAT, which is also the fused feature matrix $\boldsymbol{f}$ output from NFEN.
And $\boldsymbol{e}$ represents the edge feature matrix.

3) \textit{OAJN:}
After thoroughly extracting multi-dimensional information of plots and spatio-temporal coupling information among multi-frame plots, we employ OAJN to achieve intelligent and integrated track search and track detection process within observation association graphs. 
Compared to traditional methods that utilize energy accumulation with a given threshold for track detection \cite{ref19}, this approach excels in learning the mapping from measurement information to whether observation associations stem from the target, thus providing a better assessment of the track quality. 
First, OAJN accomplishes a link prediction task, classifying edges to obtain the confidence that the observation associations originate from the target. 
Subsequently, the confidence that the candidate tracks are derived from the target can be obtained and the track detection is performed using the final detection threshold $\gamma_2$.

Considering that $\boldsymbol{e}_{j, i}=\left[F_{E, j}, F_{E, i}\right]$ directly reflects the coupling relationship between position and Doppler measurements, we reintegrate edge features into OAJN for more effective utilization. 
Additionally, the difference in Doppler channels of the target is an important feature for target kinematic characteristics. 
For instance, the difference in Doppler channels of a target moving at a constant speed across different frames is typically small. 
As a consequence, we also incorporate the Doppler channel difference 
$d_{j, i}=\left|d_j-d_i\right|$ between node $i$ and node $j$ into OAJN to enhance the learning of link prediction. 
To facilitate the subsequent distinguishing between target tracks and false tracks based on the final detection threshold, we select the softmax function as the nonlinear activation when decoding.
In the observation association graph, edges are categorized into three types: the edge between false alarms, the edge between a false alarm and a target, and the edge between targets. 
Their corresponding desired labels are denoted as $y_{j, i}^{(1)}$, $y_{j, i}^{(2)}$, $y_{j, i}^{(3)}$. For any two connected nodes $i$ and $j$, the classification of their edge is expressed as follows:
\begin{equation}
\begin{aligned}
\hat{\boldsymbol{y}}_{j, i} & =\left[\hat{y}_{j, i}^{(1)}, \hat{y}_{j, i}^{(2)}, \hat{y}_{j, i}^{(3)}\right] \\
& =\operatorname{softmax}\left(\mathbf{W}_J^{\left(N_{J}\right)} \cdots \phi_J\left(\mathbf{W}_J^{(1)} \boldsymbol{F}_{j, i}\right)\right) \text{,} \\
\boldsymbol{F}_{j, i} & = \left( \boldsymbol{x}_i^{\left(N_G\right)} \| \hspace{0.07cm} \boldsymbol{x}_j^{\left(N_G\right)} \| 
\hspace{0.07cm} \mathbf{W}_{ed} \left( \boldsymbol{e}_{j, i} \| d_{j, i} \right) \right) \text{,}
\end{aligned}
\label{eq16}
\end{equation}
where $\hat{y}_{j, i}^{(c)}, c=1,2,3$ represents the classification score for edges, indicating the confidence that the edges originate from target associations.
$\boldsymbol{x}_{i}^{\left(N_G\right)}$ and $\boldsymbol{x}_{j}^{\left(N_G\right)}$ are the output node features of STEF-GAT and $\|$ is a concatenation operation.
$\mathbf{W}_J^\cdot$ and $\mathbf{W}_{ed}$ denote the learnable parameter matrices in OAJN.
$\boldsymbol{F}_{j, i}$ is the concatenated feature.
$N_{J}$ denotes the number of layers and $\phi_J$ is the nonlinear activation function. 
The categorical cross-entropy loss function is employed in this work, which not only alleviates the gradient vanishing problem caused by deep networks but is also equivalent to obtaining a nonlinear mapping from multi-dimensional information in raw data to target detection through maximum likelihood estimation \cite{ref48}. 
The categorical cross-entropy loss function used for integrated training is formulated as follows:
\begin{equation}
\mathcal{J}(\boldsymbol{\theta})=-\frac{1}{N_T} \sum_{n=1}^{N_T} \frac{1}{E_n} \sum_{(j, i) \in \mathcal{E}_n} \sum_{c=1}^3 y_{j, i}^{(c)} \log \hat{y}_{j, i}^{(c)}\text{,}
\end{equation}
where $N_T$ represents the number of graphs in the training set, $E_n$ denotes the number of edges in the $n$-th graph and $\mathcal{E}_n$ refers to the edge set of the $n$-th graph.

4) \textit{Track detection:}
The confidence $S$ that a candidate track originates from a target is defined as follows:
\begin{equation}
\begin{aligned}
S & =\frac{1}{N_E} \sum_{e=1}^{N_E} \rho_e+\lambda N_V \text{,}\\
\rho_e & =\alpha_1 \cdot \hat{y}_e^{(1)}+\alpha_2 \cdot \hat{y}_e^{(2)}+\alpha_3 \cdot \hat{y}_e^{(3)}\text{,}
\end{aligned}
\end{equation}
where $N_E$ represents the number of edges in the track, $N_V$ represents the track length, $\rho_e$ represents the confidence for the edge deriving from the target association, $\hat{y}_e^{(c)},c=1,2,3$ represents the classification score of the $e$-th edge, $\alpha_c$ and $\lambda$ are both hyperparameters.
For the configuration of hyperparameters, we offer some empirical recommendations. 
For a target track, the most desirable edges are those between targets, while the least desirable are those between false alarms.
Therefore, the weight for the classification score of edges between targets can be set to $\alpha_3 = 1$, while the weight for edges between false alarms can be set to $\alpha_1 = 0$.
Regarding the weight for edges between targets and false alarms, it can be set to a small positive value, such as $\alpha_2 = 0.2$.
On one hand, we aim to preserve as many edges between targets as possible, while also striving to suppress edges between targets and false alarms.
Thus, the weight should not be too large. 
On the other hand, to differentiate from edges between false alarms, and considering the presence of target observations, the weight should preferably be larger than $\alpha_1$.
Additionally, without affecting subsequent target tracking, associations between targets and false alarms can help to improve track continuity.
For example, in scenarios where a weak target is not detected in the current frame, if an association occurs between the historical target track and a false alarm without negatively impacting subsequent target tracking, it can actually improve track continuity.

The weight $\lambda$ of track length should be balanced with the weights of edges. 
From one perspective, it should not be set too high, as this may place excessive emphasis on track length, leading to an increase in false alarms within the track.
From another perspective, while ensuring that the confidence score for edges originating from target associations is high, longer tracks should be chosen to enhance track continuity. 
After obtaining the track confidence, the final detection threshold $\gamma_2$ is applied for track detection. 
Candidate tracks with the confidence score exceeding $\gamma_2$ are considered as potential target tracks. 
Finally, following track pruning, the confirmed tracks are generated and output.

\section{COMPUTATIONAL COMPLEXITY ANALYSIS}\label{C}
The computational complexity of the proposed GLP-MFD algorithm primarily consists of two components: the construction of the observation association graph and the processing of MFLPN, whose network module includes NFEN, STEF-GAT, and OAJN.
The complexity of constructing the observation association graph is the function of the observations per frame and the number of frames to be processed. 
Due to the challenges in detecting low SNR targets, it is often difficult to guarantee detection in every frame. 
Therefore, each observation needs to attempt associations with those in the subsequent $Q$ frames. 
When forming graphs from $L$ consecutive frames, for the observations in the $l$-th frame, the maximum velocity constraint is applied for association testing with observations in the set $\{l+1, \ldots, \min \{l+Q, L\}\}$.
Given that targets are typically sparse in the observation space, we explore the computational complexity under the case where only false alarms are present in each frame.
For a sensor system with $N_r$ range cells, $N_{\theta}$ beam positions and $N_D$ Doppler channels, when the primary detection threshold $\gamma_1$ is set corresponding to the initial false alarm probability $P_{fa, 1}$, the average number $N_{G C}$ of required operations for graph construction is as follows \cite{ref24}:
\begin{equation}
N_{G C}=Q\left(L-\frac{Q}{2}-\frac{1}{2}\right)\left(N_r N_{\theta} N_D P_{fa, 1}\right)^2\text{.}
\end{equation}
Let $N_{fa}$ represent the expectation of the false alarm number per frame, then we get $N_{f a}=N_r N_{\theta} N_D P_{f a, 1}$.
Consequently, the average computational complexity for graph construction can be simplified to $\mathrm{O}\left(L Q N_{f a}^2\right)$.

The operations involved in NFEN primarily consist of matrix multiplication and element-wise operations.
Its computational complexity is a function of the number of parameters. 
NFEN is divided into two main categories: the convolutional network and the fully connected networks. 
The number of convolutional layers is denoted as $N_{\boldsymbol{a}}$. 
$I_0$ and $I_{N_{\boldsymbol{a}}}$ denote the numbers of input and output channels respectively while $I_p(p=1,2, \ldots, N_{\boldsymbol{a}}-1)$ denotes the channels from intermediate convolutional operations. 
$M_p(p=1,2, \ldots, N_{\boldsymbol{a}})$ is the size of the convolutional kernel for each layer and the bias term is set to 1. 
The total number of parameters for the convolutional network is as follows:
\begin{equation}
N_{t,\boldsymbol{a}} = \sum_{p=0}^{N_{\boldsymbol{a}}-1}\left(I_p M_{p+1}+1\right) I_{p+1} \text{.}
\end{equation}
Specifically, the numbers of input and output channels are often set to 1.

\begin{figure*}
\centerline{\includegraphics[width=0.80\textwidth]{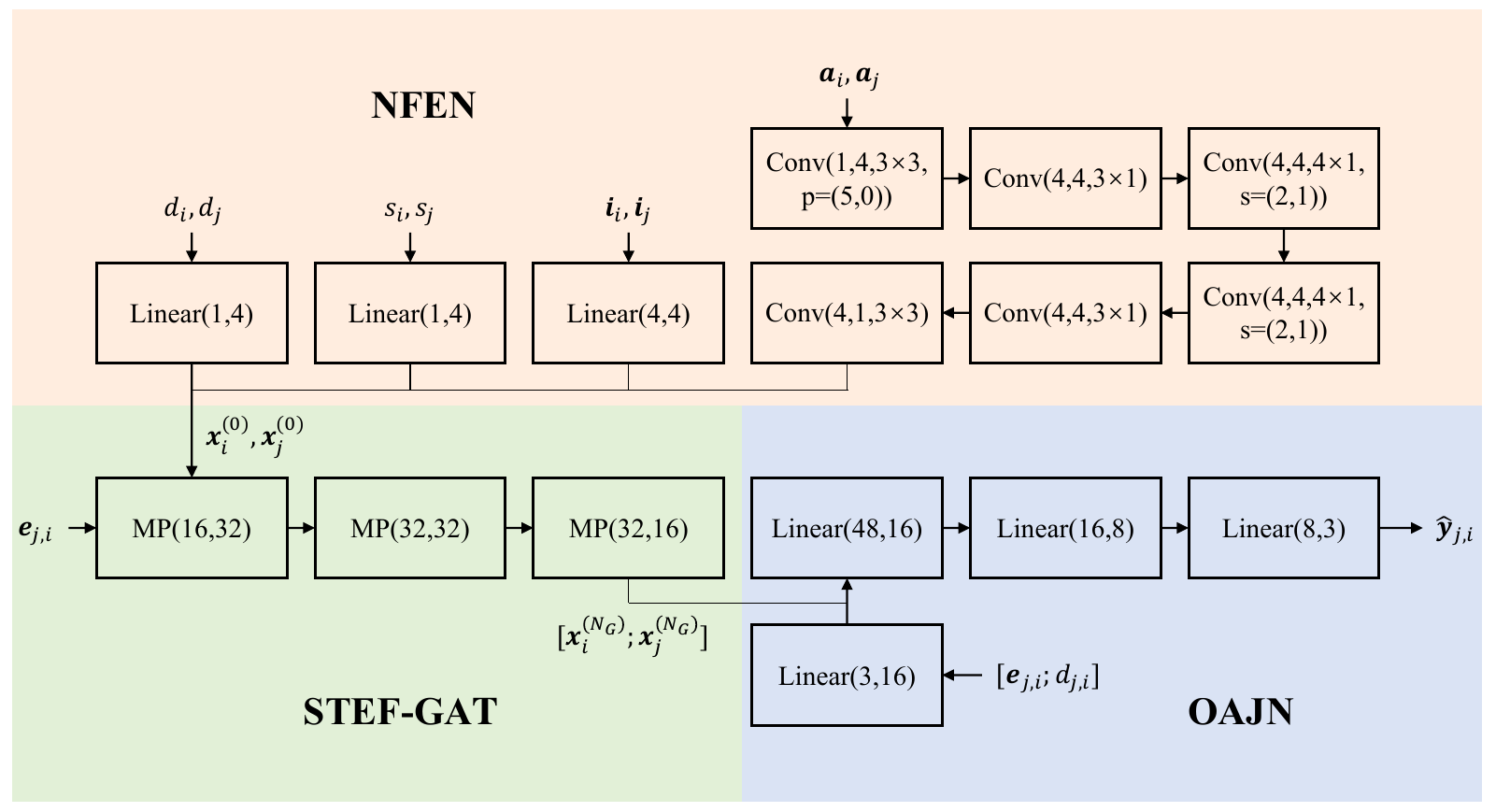}}
\caption{Detailed network structure.}
\label{fig_add_3}
\end{figure*}

To simplify the analysis, we assume that the numbers of hidden neurons in three fully connected networks are all set to $N_h$.
Let $N_d, N_s, N_{\boldsymbol{i}}$ denote the number of layers for $d_{k, n}, s_{k, n}, \boldsymbol{i}_{k, n}$, respectively. 
The total number of parameters for each of the three fully connected networks can be expressed as follows:
\begin{equation}
\begin{aligned}
N_{t, d} &= 2 N_h+\left(N_d-2\right)\left(N_h+1\right) N_h+\left(N_h+1\right) N_h \text{,}\\
N_{t, s} &= 2 N_h+\left(N_s-2\right)\left(N_h+1\right) N_h+\left(N_h+1\right) N_h \text{,}\\
N_{t, \boldsymbol{i}} &= 5 N_h+\left(N_{\boldsymbol{i}}-2\right)\left(N_h+1\right)N_h+\left(N_h+1\right) N_h \text{,}
\end{aligned}
\end{equation}
where each term in the formula corresponds to the parameters of the input layer, hidden layers, and the output layer, respectively. 
The coefficient of $N_h$ in the first term denotes the input feature dimension plus a bias which is set to 1.
Therefore, the total parameters in NFEN is:
\begin{equation}
N_{t,NFEN} = N_{t, d} + N_{t, s} + N_{t, \boldsymbol{i}} + N_{t, \boldsymbol{a}} \text{.}
\end{equation}

For STEF-GAT, the parameter number of $\mathbf{L}_E^{(k)}$ in all $N_G$ layers is $2 N_{L_E}$, where 2 is the feature dimension of $\boldsymbol{e}_{j, i}$ and $N_{L_E}$ is the dimension of the enhanced edge feature.
The parameter number of $\mathbf{W}_G^{(k)}$ is $N_{x_k} N_{x_{k+1}}$, where $N_{x_k}$ represents the feature dimension of nodes in the $k$-th layer.
$\beta_G$ is a fully connected layer with $\left(N_{L_E}+N_{x_{k+1}}\right) N_{x_{k+1}}$ parameters. 
The parameter number of $\mathbf{W}_E^{(k)}$ in all $N_G$ layers is $2 N_{W_E}$, where $N_{W_E}$ is the dimension of the output edge feature in the attention mechanism.
The parameter number of $\boldsymbol{a}_G^{(k)}$ is $\left(N_{W_E}+2 N_{x_{k+1}}\right)$.
Therefore, the total number of parameters in STEF-GAT is as follows:
\begin{equation}
\begin{aligned}
 N_{t,STEF} = & \sum_{k=0}^{N_G-1}\left[2 N_{L_E}+N_{x_k} N_{x_{k+1}}+3 N_{W_E}+\right. \\
& \left.\left(N_{L_E}+N_{x_{k+1}}\right) N_{x_{k+1}}+2 N_{x_{k+1}}\right]
\end{aligned}
\end{equation}

As to OAJN, let $N_m$ and $N_J$ denote the number of hidden neurons and the number of layers with the bias term set to 1 , respectively. 
The total number of parameters in OAJN according to \eqref{eq16} is as follows:
\begin{equation}
\begin{aligned}
N_{t, O A J N} & = \left(2+1+1\right) N_m + \left(2 N_{x_{N_G}}+N_m+1\right) N_m \\
& +\left(N_J-2\right)\left(N_m+1\right) N_m +3\left(N_m+1\right) \text{.}
\end{aligned}
\end{equation}
In summary, when performing integrated track search and track detection in an observation association graph, the computational complexity of MFLPN can be expressed as follows:
\begin{equation}
\begin{aligned}
\mathrm{O} &\left(N_{\boldsymbol{a}} M_{\max } I_{\max }^2+\left(N_d+N_s+N_{\boldsymbol{i}}\right) N_h^2\right. \\
& \left.+N_G N_{x, \max }^2+N_J N_m^2\right)\text{,}
\end{aligned}
\end{equation}
where $M_{\max }$ and $I_{\max }$ represent the maximum kernel size and the maximum number of channels in NFEN, respectively.
And $N_{x, \max }$ represents the maximum feature dimension of nodes in STEF-GAT.

In practical engineering applications, the entire observation space can typically be divided into $N_{\tau}$ subareas.
Tracks are searched and detected within each subarea to reduce the computational burden on the system.
Therefore, the computational complexity of the graph construction will decrease to $\mathrm{O}\left(L Q N_{f a}^2 / N_{\tau}\right)$.
In addition, parallelizing the initiation of tracks in subareas should enhance computational performance. 
The number of parameters in the network determines its computational complexity, allowing for optimization such as pruning the parameters that have minimal impact on the network. 
Moreover, GPU parallel processing can be utilized for operations like matrix multiplication to further improve computational efficiency of the proposed algorithm.

\section{NUMERICAL RESULTS}\label{D}
In this section, numerical experiments are conducted to show the performance improvement in multi-frame detection achieved by the proposed GLP-MFD.
The primary threshold $\gamma_1$ determines the optimal detection performance that the proposed algorithm can achieve. 
Therefore, to ensure effective detection performance for weak targets while avoiding excessive false alarms that will impose a computational burden on the system, we set $\gamma_1$ corresponding to an initial false alarm probability $P_{f a, 1}=10^{-3}$.
Additionally, it is assumed that false alarms originate from internal receiver noise, which is generally valid for long-range target detection.

When processing multi-frame data using neural networks, the relationship among $\gamma_2$, the final false alarm probability, and the detection probability is complex. 
Consequently, the final false alarm probability and detection probability of the proposed algorithm are calculated using Monte Carlo experiments in this work.
Since GLP-MFD transforms plot-level detection into track-level detection, the final false alarm probability $P_{f a, 2}$ based on $\gamma_2$ is defined as the total number of false tracks divided by the total number of detection cells. 
The detection probability is defined as the total number of correctly detected target tracks divided by the total number of target appearances in the Monte Carlo experiments.

\begin{figure*}
\centerline{\includegraphics[width=0.82\textwidth]{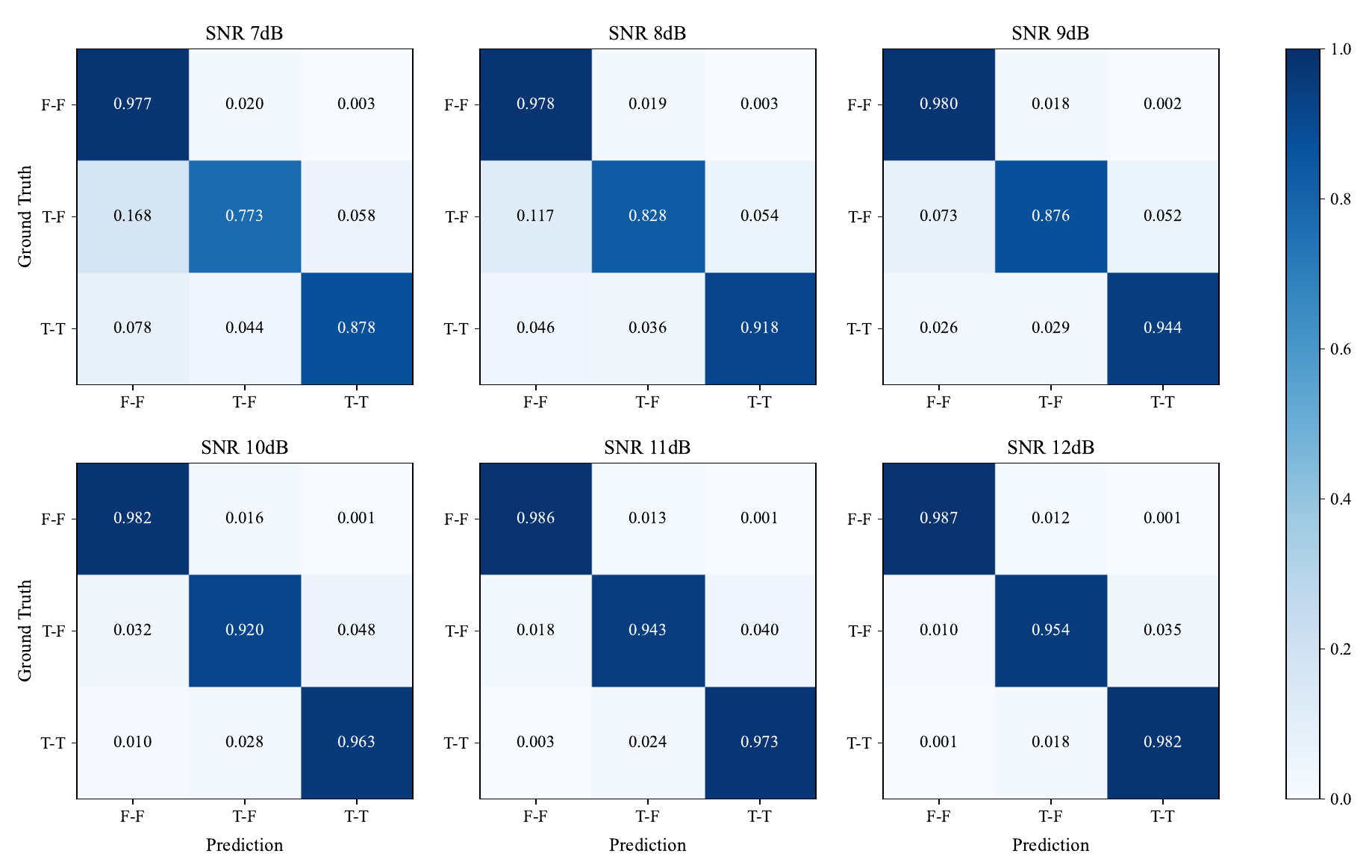}}
\caption{Confusion matrices for different SNRs.}
\label{fig_add_2}
\end{figure*}

\subsection{Network performance analysis}
The proposed MFLPN performs a link prediction task in observation association graphs and is implemented using the PyTorch framework \cite{ref_add_2}.
The dataset is simulated based on an L-band radar with $2^{\circ}$ azimuth resolution (3dB antenna), 100 m range resolution and 32 pulses transmitted at each beam position. 
The graph used for training is constructed by multi-frame observations that exceed the primary threshold, based on the method described in section \ref{B}.\ref{subsec:3A}.
The hardware configuration is NVIDIA GeForce RTX 4090 GPU.
We set the batch size to 32 and use the Adam optimizer for training the network \cite{ref_add_3}.
The total number of training epochs is 800 and the initial learning rate is set to 0.01, which is reduced to 1/10 of its previous value every 200 epochs. 
To show the key hyperparameters in the proposed network, we provide the detailed network structure as shown in Fig \ref{fig_add_3}. 
For better readability, we take the link prediction between node $i$ and node $j$ in a graph as an example.
The description ``Conv(4,1,3$\times$3)" signifies a convolutional layer with 4 input channels, 1 output channel and a 3$\times$3 kernel size.
``p=(5,0)" refers to a padding of 5 in height and 0 in width, with a circular padding mode. 
If description ``p" is not provided, no padding operation is performed.
``s=(2,1)" refers to a stride of 2 in height and 1 in width. 
If ``s" is not provided, the default stride is 1 in both height and width.
``Linear(1,4)" denotes a fully connected layer with an input dimension of 1 and an output dimension of 4.
``MP(16,32)" denotes a message passing layer with an input dimension of 16 and an output dimension of 32.
The meaning of the remaining descriptions can be inferred by analogy.

Link prediction can be regarded as a classification task for edges in the graph. 
In this work, the edges are classified into three categories: false alarm to false alarm (F-F), target to false alarm (T-F), and target to target (T-T). 
To demonstrate the network performance, we present the confusion matrices under different SNR conditions, as shown in Fig. \ref{fig_add_2}.
It can be observed that targets with higher SNR, due to their more distinct target characteristic, result in better classification performance.
In addition, by observing the network accuracy for each structure, we conduct the ablation study to validate the effectiveness of each individual module in the proposed network.
Moreover, to further demonstrate the advantage of message passing in STEF-GAT, we choose the Graph Convolutional Network (GCN) for comparison \cite{ref_add_4}.
GCN is a classical method in graph neural networks and we replace STEF-GAT with GCN in the full network.
The network accuracy for various SNRs in the ablation study is shown in Table \ref{tab2}.
The result indicates that the designed NFEN, STEF-GAT and OAJN all contribute positively to the overall network performance. 
NFEN has the ability to effectively pre-extract node features, aiding the network training process.
OAJN facilitates better utilization of the extracted features to accomplish the required task.
STEF-GAT, as the most crucial module in the network, is capable of learning the spatio-temporal coupling between multiple frames, further extracting node features with multi-frame fusion. 
It maximally supports the proposed algorithm to perform integrated track search and track detection in observation association graphs.
Since the message passing in GCN fails to incorporate the attention mechanism and does not design edge features based on spatio-temporal coupling information, the contribution of GCN is far inferior to that of STEF-GAT.
This comparison further highlights the superiority of the designed message passing in STEF-GAT.

\begin{table*}[htbp]
    \centering
	\renewcommand\arraystretch{1.25}
    \setlength{\tabcolsep}{12pt}
	\caption{Ablation study result.}
    \begin{tabular}{>{\centering\arraybackslash}p{4cm}@{\hspace{8pt}}cccccc}
        \hline
        Structure & 7dB & 8dB & 9dB & 10dB & 11dB & 12dB \\
        \hline
        \raggedright \hspace{9pt}NFEN        &68.19\%  &70.30\%  &74.24\%  &77.23\%  &80.70\%  &82.37\%  \\
        \raggedright \hspace{9pt}STEF-GAT    &71.48\%  &76.50\%  &80.52\%  &85.60\%  &89.04\%  &90.68\%  \\
        \raggedright \hspace{9pt}OAJN        &64.71\%  &67.32\%  &70.65\%  &73.94\%  &77.41\%  &79.36\%  \\
        \raggedright \hspace{9pt}NFEN + STEF-GAT    &81.18\%  &85.02\%  &88.63\%  &92.26\%  &94.02\%  &95.32\%  \\
        \raggedright \hspace{9pt}NFEN + OAJN        &72.94\%  &75.24\%  &78.14\%  &81.04\%  &84.13\%  &85.73\%  \\
        \raggedright \hspace{9pt}STEF-GAT + OAJN    &77.71\%  &82.47\%  &86.19\%  &89.60\%  &92.46\%  &93.77\%  \\
        \raggedright \hspace{9pt}NFEN + GCN + OAJN            &74.36\%  &78.20\%  &80.46\%  &84.26\%  &85.97\%  &87.22\%  \\
        \raggedright \hspace{9pt}NFEN + STEF-GAT + OAJN       &87.60\%  &90.84\%  &93.33\%  &95.51\%  &96.73\%  &97.44\%  \\
        \hline
    \end{tabular}
    \label{tab2}
\end{table*}
\begin{table}[htbp]
    \centering
	\renewcommand\arraystretch{1.25}
    \setlength{\tabcolsep}{4.5pt}
	\caption{Network accuracy with different layers in STEF-GAT.}
    \begin{tabular}{c|ccccc}
        \hline
        Layer number       &1  &2  &3  &4  &5   \\
        \hline
        Accuracy  &89.55\%  &91.15\%  &93.58\%  &91.64\%  &91.87\%  \\
        \hline
    \end{tabular}
    \label{tab_add_1}
\end{table}
\begin{table}[htbp]
    \centering
    \renewcommand\arraystretch{1.55}
    \setlength{\tabcolsep}{6pt}
    \caption{Runtimes of the network.}
    \begin{tabular}{c@{\hskip 4pt}c@{\hskip 5pt}|c@{\hskip 6.5pt}c@{\hskip 6.5pt}c@{\hskip 6.5pt}c}
        \hline
        \multicolumn{2}{c|}{Node Number} & $10^2$ & $10^3$ & $10^4$ & $10^5$ \\
        \hline
        \multirow{2}{*}{\parbox{1.6cm}{\raggedright\vspace{3pt}\hspace{7pt}NFEN \\ + STEF-GAT \\ + OAJN\vspace{3pt}}}
                                  & GPU  &0.0069 s  &0.0082 s  &0.0172 s  &0.1644 s\\
                                  & CPU  &0.0085 s  &0.0443 s  &0.2439 s  &1.5588 s\\
        \hline
        \multirow{2}{*}{\parbox{1.6cm}{\raggedright\vspace{3pt}\hspace{7pt}NFEN \\ + GCN \\ + OAJN\vspace{3pt}}} 
                                  & GPU  &0.0053 s  &0.0058 s  &0.0108 s  &0.1322 s\\
                                  & CPU  &0.0054 s  &0.0292 s  &0.1626 s  &1.0801 s\\
        \hline
    \end{tabular}
    \label{tab3}
\end{table}

The configuration of hyperparameters should take into account both practical requirements and physical significance. 
Given that the radial velocity of the moving target varies across multiple frames, it is crucial to ensure that the receptive field of the convolutional neural network in NFEN can cover the changes in the target Doppler channels.
The fully connected networks in NFEN aim to map input features to a higher dimensional space and the purpose of OAJN is to better perform link prediction based on the features extracted by previous modules.
The multi-layer message passing in STEF-GAT implies the fusion of multi-frame information.
However, it is difficult to ensure successful detection of weak targets in each frame.
Therefore, it is necessary to balance the number of frames fused for message passing of multi-frame target observations with the increased training difficulty and computational cost brought by excessively deep layers. 
Methods such as grid search, random search, and Bayesian optimization can be used to jointly tune the hyperparameters \cite{ref_hyper}.
When the hyperparameter search space is large, Bayesian optimization can significantly reduce the search iterations and more efficiently identify a hyperparameter configuration that is close to the optimum.
Bayesian optimization commonly involves the following steps.
First, an objective to be optimized and the input hyperparameter space are defined.
Next, a surrogate model like Gaussian process is selected to fit the objective function.
Then, an acquisition function like expected improvement is used to sample the hyperparameter space and evaluate the objective.
Finally, the surrogate model is iteratively updated based on the evaluation results.
When tuning the hyperparameters of the proposed network, the network accuracy can be chosen as the objective to be optimized, and the hyperparameter search space can be defined based on each module. 
For example, the search range for the number of convolutional layers in NFEN can be set between 1 and 8, while the range for the number of layers in STEF-GAT and OAJN can be set between 1 and 5. 
The number of hidden neurons can range from 8 to 64, while the learning rate and batch size can range from $10^{-1}$ to $10^{-5}$ and 16 to 64, respectively. 
The tool scikit-optimize can be used for hyperparameter optimization of the proposed network.
In addition, to further demonstrate the impact of the number of layers in the key model STEF-GAT on network performance, we provide the average network accuracy across various SNRs for different layer configurations during message passing, as shown in Table \ref{tab_add_1}.
The best performance occurs with 3 layers of message passing. 
When the number of layers is small, the accuracy may decline due to insufficient fusion of multi-frame information.
Conversely, when the number of layers is large, the accuracy may degrade due to increased training difficulty, such as overfitting.

To validate the time cost of the proposed network, we calculate the average runtime for an observation association graph using the Monte Carlo experiment. 
The runtime of a graph is directly related to its scale.
As the graph scale increases, the time cost will correspondingly rise. 
The graph scale is closely related to factors such as the size of the divided subareas, the initial false alarm probability for the primary detection threshold, the number of processing frames, and the number of subareas processed in parallel. 
For a more intuitive demonstration, we reflect the graph scale by the number of nodes in the graph.
For example, in a subarea with $10^5$ detection cells, where the initial false alarm probability is $10^{-3}$, with $5$ processing frames and $20$ subareas processed in parallel, the number of nodes in the graph is approximately $10^4$. 
As shown in Table \ref{tab3}, we provide the runtime for different graph scales on Intel(R) Xeon(R) Platinum 8370C CPU and NVIDIA GeForce RTX 4090 GPU.
To further validate the computational efficiency of the key module STEF-GAT, we replace it with GCN in the full network.
Therefore, we test the runtime of the network consisting of NFEN, STEF-GAT, and OAJN, as well as the network consisting of NFEN, GCN, and OAJN.
It can be seen that for a network including a large number of matrix multiplications and element-wise operations, GPU has a significant advantage in computational efficiency compared to CPU.
When the graph scale is relatively large, this advantage becomes more prominent.
Since GCN has fewer parameters, it takes less time to run. 
The runtime of the network including STEF-GAT is on the same order of magnitude as the network including GCN, indicating that STEF-GAT does not bring a notable computational cost.
However, as shown in Table \ref{tab2}, the contribution of STEF-GAT to the improvement in network accuracy is significant compared to GCN.

\subsection{The definition of correct target detection}
The proposed GLP-MFD enables direct output of target tracks through the link prediction task in observation association graphs. 
To evaluate the performance of GLP-MFD, it is necessary to define whether the output track correspond to a correct target track. 
Due to measurement noise and the dense false alarm environment, it is possible for a false alarm to exhibit a quality superior to that of a true target measurement, which may lead to the inclusion of false alarms in the initiated target track.
However, if the existence of false alarms does not affect subsequent target tracking, the initiated target track can still be considered as a correct target track.
In light of the points mentioned above, we employ the optimal subpattern assignment (OSPA) distance \cite{ref49} to determine whether the output target track has been correctly detected. 
The OSPA distance is a metric used in target tracking tasks to measure the difference between the set of estimated target states and true target states at a given time, by defining a distance measure in the state space.
Given that the OSPA distance is essentially a metric for measuring the discrepancy between sets, in the context of this work, we use it to quantify the difference between the smoothed confirmed target tracks output from MFLPN and the tracks with correct target states.
In this manner, we determine whether the output target tracks are indeed correct.
Techniques such as linear regression, quadratic regression, and Kalman filtering can be utilized as smoothing algorithms.
The formula for the OSPA distance in the context of our work is as follows:
\begin{equation}
\begin{aligned}
& d_{\xi}^{(\eta)}(\boldsymbol{X}, \hat{\boldsymbol{X}})=\left(\frac{1}{N_{\hat{\boldsymbol{X}}}} \min _{\pi \in \Pi_{N_{\hat{\boldsymbol{X}}}}} \sum_{i=1}^{N_{\boldsymbol{X}}} d^{(\eta)}\left(\boldsymbol{x}_i, \hat{\boldsymbol{x}}_{\pi(i)}\right)^{\xi}\right. \\
& \hspace{1.73cm} \left.+\frac{\eta^{\xi}}{N_{\hat{\boldsymbol{X}}}}\left(N_{\hat{\boldsymbol{X}}}-N_{\boldsymbol{X}}\right)\right)^{1 / \xi} , N_{\boldsymbol{X}} \leq N_{\hat{\boldsymbol{X}}}  \text{,} \\
& d_{\xi}^{(\eta)}(\boldsymbol{X}, \hat{\boldsymbol{X}})=d_{\xi}^{(\eta)}(\hat{\boldsymbol{X}}, \boldsymbol{X}) \hspace{1.48cm} , N_{\boldsymbol{X}} > N_{\hat{\boldsymbol{X}}} \text{,}\\
& d^{(\eta)}\left(\boldsymbol{x}_i, \hat{\boldsymbol{x}}_{\pi(i)}\right)=\min \left(\eta, d\left(\boldsymbol{x}_i, \hat{\boldsymbol{x}}_{\pi(i)}\right)\right)\text{,}
\end{aligned}
\end{equation}
where $\hat{\boldsymbol{X}}=\left(\hat{\boldsymbol{x}}_1, \hat{\boldsymbol{x}}_2, \ldots, \hat{\boldsymbol{x}}_{N_{\hat{\boldsymbol{X}}}}\right)$ represents the set of smoothed confirmed target tracks output from MFLPN, $\boldsymbol{X}=\left(\boldsymbol{x}_1, \boldsymbol{x}_2, \ldots, \boldsymbol{x}_{N_{\boldsymbol{X}}}\right)$ represents the set of tracks with correct target states, 
$\hat{\boldsymbol{x}}_i$ and $\boldsymbol{x}_i$ represent the plot position in tracks, 
$N_{\hat{\boldsymbol{X}}}$ and $N_{\boldsymbol{X}}$ represent the total plot number of tracks, $d\left(\boldsymbol{x}_i, \hat{\boldsymbol{x}}_{\pi(i)}\right)$ represents the Euclidean distance between $\boldsymbol{x}_i$ and $\hat{\boldsymbol{x}}_{\pi(i)}$,
$\Pi_{N_{\hat{\boldsymbol{X}}}}$ represents the set of permutations on $\left\{1,2, \ldots, N_{\hat{\boldsymbol{X}}}\right\}$,
$\xi$ represents the order parameter and $\eta$ represents the cutoff parameter.

$\xi$ affects the deviation between the estimation and the true value. 
As $\xi$ increases, the weight of $d_{\xi}^{(\eta)}(\boldsymbol{X}, \hat{\boldsymbol{X}})$ relative to the position error enhances.
$\eta$ influences the discrepancy in the number of plots between smoothed tracks and correct tracks.
As $\eta$ increases, the weight of $d_{\xi}^{(\eta)}(\boldsymbol{X}, \hat{\boldsymbol{X}})$ with respect to the mismatch in the number of plots enhances.
In the context of our work, we focus on the impact of false alarms on the quality of target tracks, with greater emphasis placed on position errors. 
Considering practicality and the smoothness of distance curves, we set $\xi = 2$ \cite{ref50}.
Moreover, the value of $\eta$ should be relatively small. 
After evaluating the position measurement errors in terms of the variance of measurement noise, the value of $\eta$ determined based on the position measurement errors can be regarded as a small value \cite{ref50}.
Given that the range of $d_{\xi}^{(\eta)}(\boldsymbol{X}, \hat{\boldsymbol{X}})$ is $[0, \eta]$, 
when the OSPA distance between smoothed tracks and tracks with correct target states meets the following condition:
\begin{equation}
d_{\xi}^{(\eta)}(\boldsymbol{X}, \hat{\boldsymbol{X}})<\kappa \eta \text{,}
\label{eq27}
\end{equation}
the confirmed tracks output from the network can be considered as correct target tracks.
In \eqref{eq27}, the configuration of $\kappa$ needs to ensure that the trajectories of perceived correct target tracks output from the network are approximately consistent with the tracks with correct target states, without negatively impacting subsequent tracking. 
This can be achieved by performing Monte Carlo experiments to statistically evaluate the OSPA distance between smoothed tracks consisting entirely of target measurements and the tracks with correct target states, using this evaluation as a reference for setting $\kappa$.

\begin{figure*}
\centerline{\includegraphics[width=0.87\textwidth]{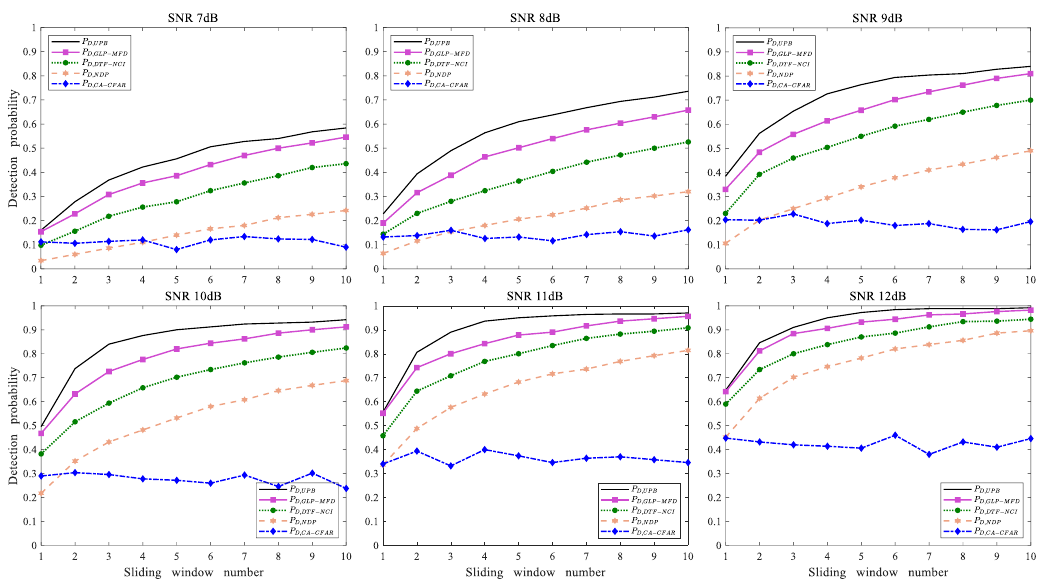}}
\caption{The detection probability versus the sliding window number.}
\label{fig5}
\end{figure*}

\begin{figure*}[ht]
    \centering
    \begin{subfigure}[b]{0.41\textwidth}
        \centering
        \includegraphics[width=\textwidth]{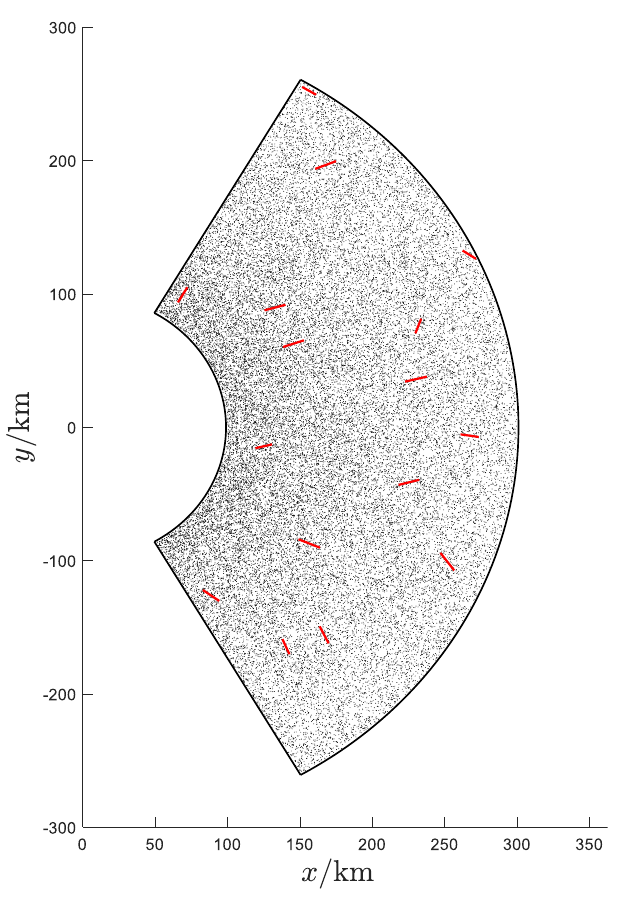} 
        \caption{Detection result of the primary threshold.}
        \label{fig6a}
    \end{subfigure}
    \hspace{-0.15cm} 
    \begin{subfigure}[b]{0.41\textwidth}
        \centering
        \includegraphics[width=\textwidth]{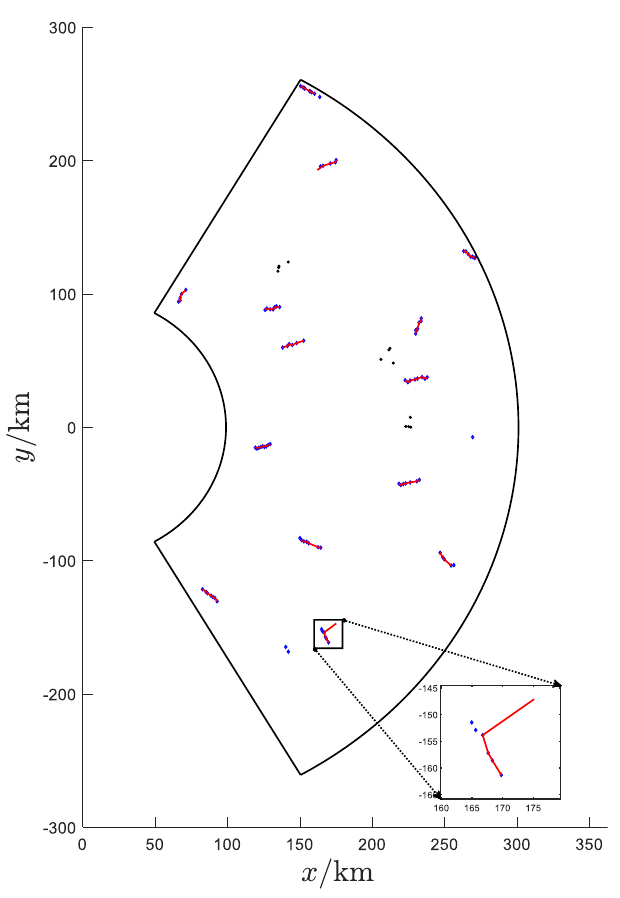} 
        \caption{Processing result of the proposed algorithm.}
        \label{fig6b}
    \end{subfigure}
    \caption{Results for one exemplar run.}
    \label{fig6}
\end{figure*}

\subsection{The detection performance analysis}
To demonstrate the superiority of the proposed GLP-MFD, a comparison is conducted with the three-stage MFD method using the novel dynamic programming (NDP)\cite{ref19}, and the method that combines Doppler-aided track formation with non-coherent integration (DTF-NCI)\cite{ref24}.
The NDP method constructs potential tracks using kinematic constraints derived from target position measurements.
And it makes track detection through non-coherent integration of track energy followed by a comparison with a final detection threshold.
The DTF-NCI method designs kinematic constraints based on Doppler information to obtain potential tracks. 
By incorporating Doppler information, the separability between targets and false alarms is enhanced.
Subsequently, the method accumulates track energy incoherently and makes track detection based on a final detection threshold.
It is important to note that the NDP method does not utilize Doppler information. 
In the experimental scenario set for this work, if the NDP method uses the primary threshold corresponding to the initial false alarm probability $P_{f a, 1}=10^{-3}$, it will lead to a combinatorial explosion, resulting in an unmanageable computational load. 
Additionally, due to its poor separability between targets and false alarms, an excessive number of false alarms will adversely impact correct associations of target measurements, degrading the performance of target detection. 
Therefore, for the NDP method, we employ the primary threshold corresponding to the initial false alarm probability $P_{f a, 1, N D P}=10^{-4}$.
The proposed GLP-MFD method and the DTF-NCI method both utilize the primary threshold corresponding to $P_{f a, 1}=10^{-3}$.
To ensure a fair comparison, all methods guarantee that the final false alarm probability $P_{f a, 2}$ have to be $10^{-6}$ after processing multi-frame data. 

Under the same SNR conditions, the relationship between target detection probabilities and sliding window numbers for aforementioned methods is shown in Fig. \ref{fig5}. 
$P_{D, C A-C F A R}$ represents the target detection probability in a single frame using the cell-averaging constant false alarm rate (CA-CFAR) method under a false alarm probability of $10^{-6}$.
$P_{D, UPB}$ denotes the upper limit of detection performance, defined as the probability that a candidate track exists where all plots are target observations across multiple frames.
It can be observed from Fig. \ref{fig5} that as the sliding window number increases, the detection probability of multi-frame detection methods gradually improves. 
In most cases, multi-frame detection methods achieve better performance compared to CA-CFAR.
However, the detection performance of NDP is mostly lower than that of CA-CFAR when the number of sliding windows is small.
On one hand, the NDP method does not utilize Doppler information, resulting in poor distinction between targets and false alarms.
In addition, the exclusion of false tracks mainly relies on energy accumulation. 
On the other hand, the primary threshold is set corresponding to $P_{f a, 1, N D P}=10^{-4}$, which restricts the upper limit of detection performance.
In contrast, the DTF-NCI method, which incorporates Doppler information, demonstrates better separability between targets and false alarms, outperforming NDP. 
However, weak targets and the dense false alarm environment can diminish the separability.
Consequently, when the SNR is 7 dB, the detection performance of DTF-NCI with a single sliding window is slightly lower than that of CA-CFAR.
Whereas, under high SNR conditions, its performance consistently surpasses CA-CFAR.

\begin{figure*}[ht]
    \centering
    \begin{subfigure}[b]{0.68\textwidth}
        \centering
        \includegraphics[width=\textwidth]{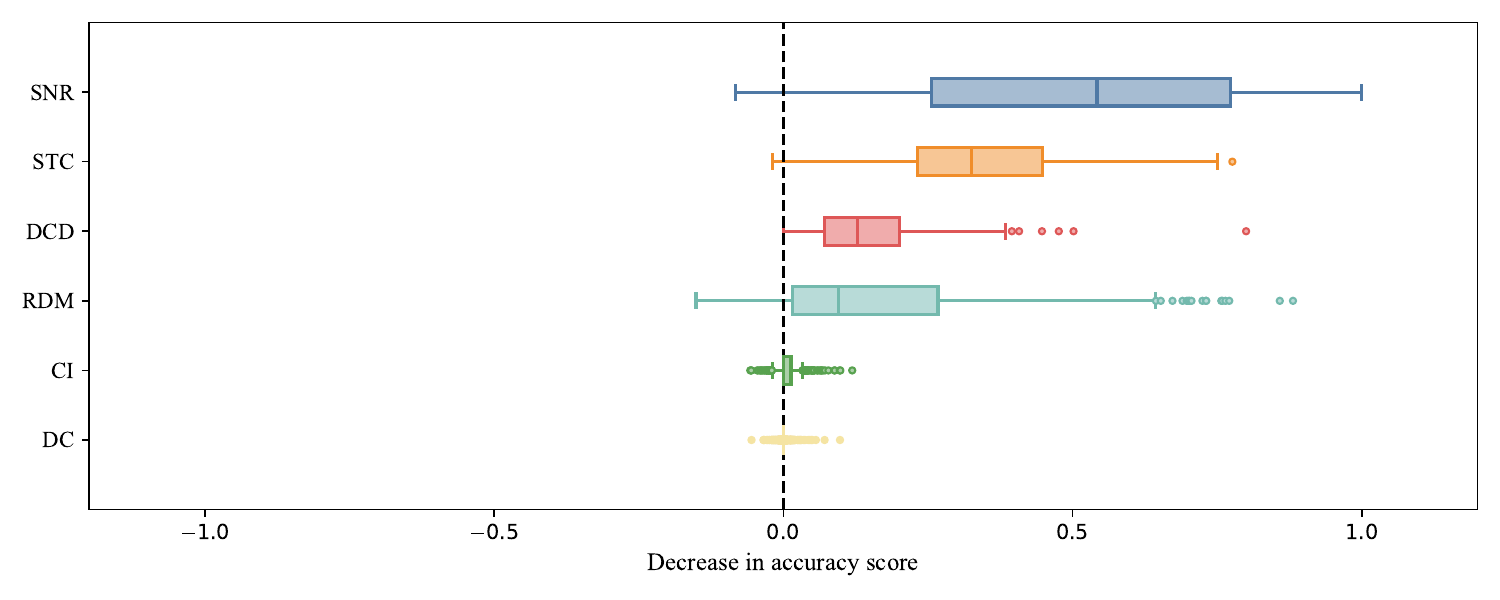}
        \caption{Boxplot of feature importance during target association decision at high SNR.}
        \label{fig7a}
    \end{subfigure}
    
    \begin{subfigure}[b]{0.68\textwidth}
        \centering
        \includegraphics[width=\textwidth]{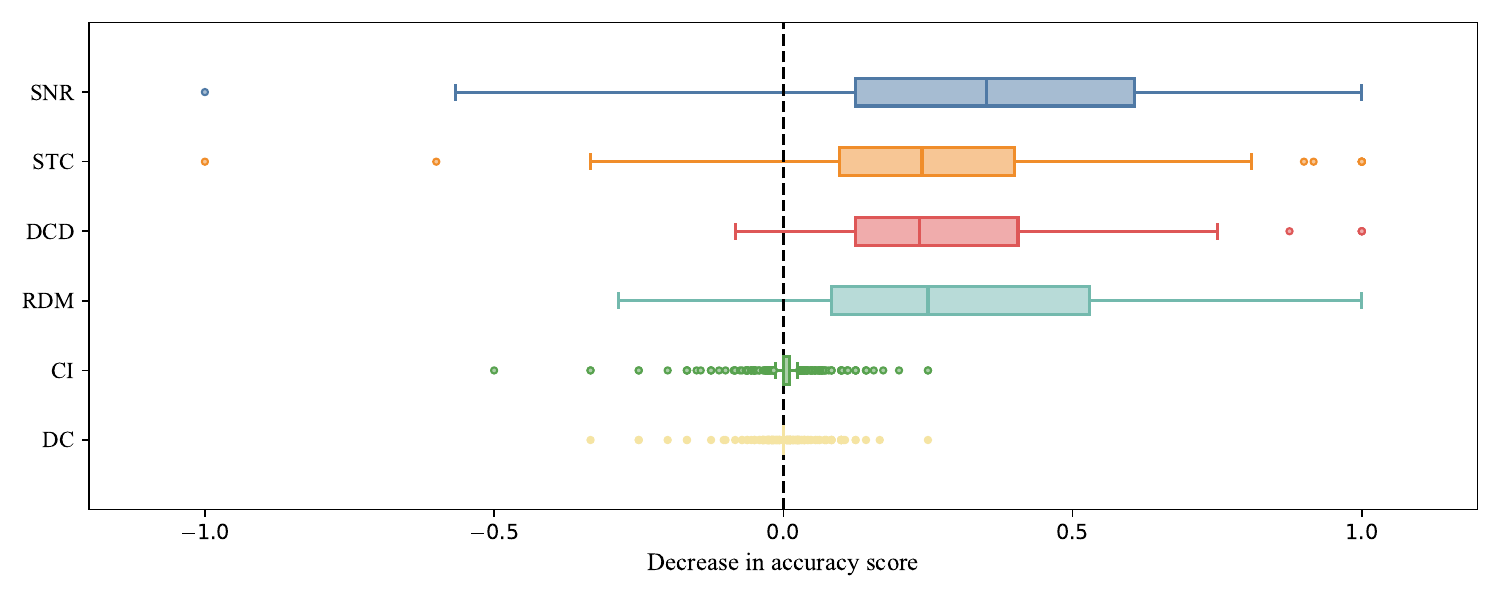}
        \caption{Boxplot of feature importance during target association decision at low SNR.}
        \label{fig7b}
    \end{subfigure}
    
    \begin{subfigure}[b]{0.68\textwidth}
        \centering
        \includegraphics[width=\textwidth]{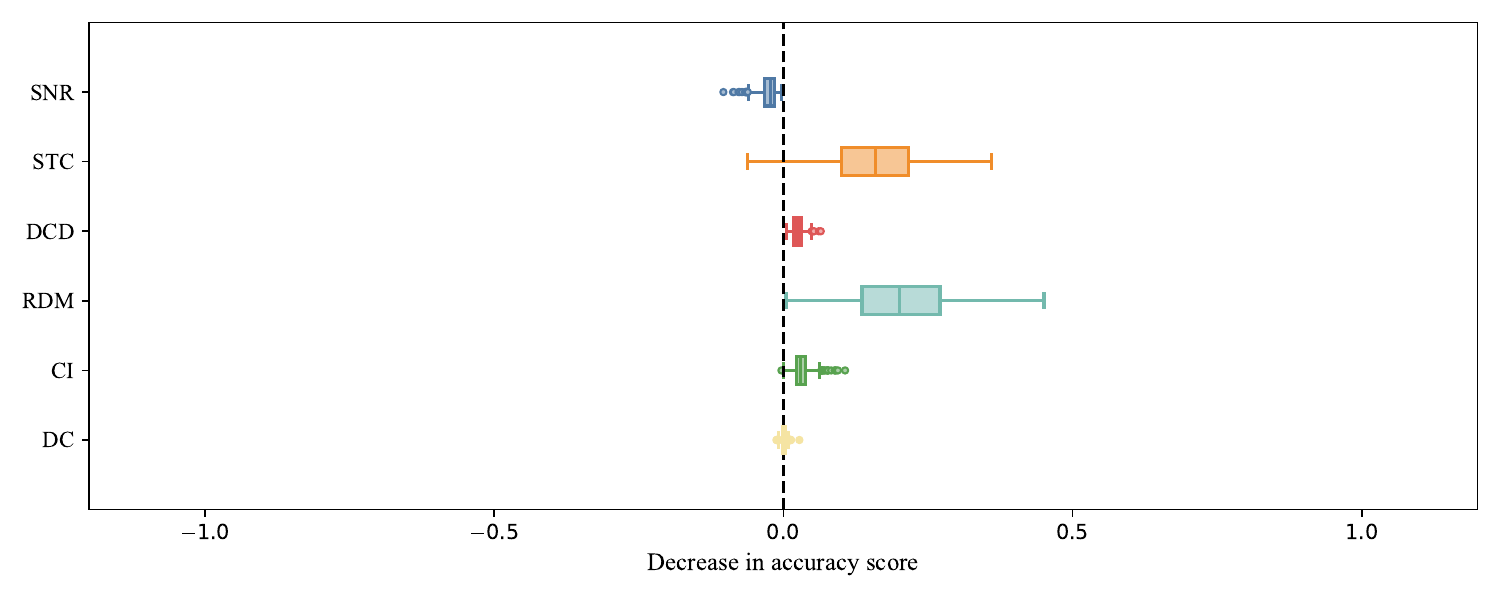}
        \caption{Boxplot of feature importance during false alarm association decision.}
        \label{fig7c}
    \end{subfigure}
    
    \caption{Boxplots of feature importance during various association decisions.}
    \label{fig7}
\end{figure*}

The CA-CFAR method exhibits poor detection performance for weak targets because it only uses single-frame information.
Although the NDP method utilizes multi-frame information for target detection, it merely uses position measurements when searching potential tracks and does not fully utilize Doppler information, which reduces the separability between targets and false alarms. 
The DTF-NCI method incorporates Doppler information to construct kinematic constraints when searching potential tracks, but such relatively strict physical constraints may lead to target missed associations, resulting in performance loss. 
Additionally, the non-coherent integration track detector does not fully exploit the multi-dimensional information on different attributes across frames, which leads to a performance bottleneck \cite{ref_distr}.
The proposed GLP-MFD algorithm integrates multi-dimensional information of target observations and spatio-temporal coupling information across multiple frames, approaching a detector for fusing features on different attributes and achieving the highest detection performance. 
Moreover, its superiority becomes more apparent at low SNR conditions compared with NDP and DTF-NCI.
It is important to note that in Monte Carlo experiments, situations may arise where a target is not detected by $P_{D, UPB}$, yet the proposed algorithm concludes that the target has been correctly detected. 
This is because GLP-MFD defines correct detection based on the OSPA distance, which may occasionally obtain the output track where false alarms are present and the number of target observations is less than $M$. 
According to the definition of $P_{D, UPB}$, this does not count as a correct detection.
However, the compliance of the condition shown in \eqref{eq27} indicates that the false alarms in the track do not affect subsequent target tracking, leading the consideration of correct detection. 
In practical engineering applications, such high-quality false alarms can also be advantageous for target track continuity.

To completely show the experimental scenario and the performance of GLP-MFD, the processing results for one exemplar run are shown in Fig. \ref{fig6}.
The observation area of interest is from $100$ km to $300$ km and from $-60^{\circ}$ to $60^{\circ}$ containing a total of 16 targets. 
The detection result corresponding to the primary threshold is shown in Fig. \ref{fig6a}, with red solid lines indicating the tracks with correct target states. 
The processing results of the proposed GLP-MFD is displayed in Fig. \ref{fig6b}, where the red solid lines represent the target tracks output from the network, black solid points indicate the false tracks, and blue squares denote the position measurements of targets.

In the processing result, 3 false tracks and 14 target tracks are detected, while two target tracks are not detected. 
The retention of false tracks may result from their alignment with the multi-frame correlation characteristics of targets.
However, they can be further eliminated through false track discrimination methods.
Among 14 detected target tracks, 13 are identified as correct target tracks. 
The enlarged track in the figure has a large OSPA distance from the true target track due to the presence of false alarms, which affects the target track direction and may impact subsequent tracking. 
Therefore, it cannot be considered correctly detected.
Nevertheless, these erroneous target tracks do not impose excessive negative effects. 
Over continuous multi-frame processing, the incorrect detection will gradually be filtered out, while the correct target observations within the erroneous tracks will remain and appear in the newly detected correct tracks. 
In summary, the proposed GLP-MFD algorithm demonstrates better detection performance for weak targets and has the ability to accurately detect target tracks while effectively suppressing false tracks.

\subsection{Interpretability analysis}
To explore how MFLPN performs integrated track search and track detection based on input features, we analyze the importance of each input feature on the network output. 
The input features include signal-to-noise ratio (SNR), Doppler channel (DC), chronological information (CI), range-Doppler map (RDM), spatio-temporal coupling (STC) and the differences in Doppler channels (DCD).
To analyze the contribution of these features to the network, techniques such as feature occlusion and feature permutation can be employed to eliminate the dependency between samples and their corresponding features \cite{ref51},\cite{ref52},\cite{ref53}.
The decline in network performance can be observed to confirm the importance of each feature. 
The core task of this work is the link prediction task, which determines whether the associations between observations originate from targets. 
To clarify the network decision criteria under different conditions, we use the extent of accuracy decline in link predictions caused by eliminating sample-feature dependencies as a reference metric. 
The box plots are used to explain the feature importance for the network \cite{ref54}.
The results of association decision in high SNR targets, low SNR targets, and false alarms are shown in Fig. \ref{fig7}.

The results indicate that for high SNR targets, SNR is the most significant feature influencing performance. 
This is because a higher SNR enhances the separability between targets and false alarms, making it a key characteristic for distinguishing them. 
The next important features are the spatio-temporal coupling and the differences in Doppler channels.
Effectively utilizing these features can enhance the accuracy of target detection.
For low SNR targets, although SNR remains the most influential feature, its impact reduces compared to high SNR targets.
To compensate, the network increasingly relies on the range-Doppler map, which reflects not only the intensity of echo signals, but also the distribution of target echo energy across the range and Doppler dimensions. 
This signal structure information facilitates deeper feature extraction by the network. 
Furthermore, since the network accuracy in making association decisions for low SNR targets is lower than that for high SNR targets, the range of outliers tends to be larger.
In the case of associations between false alarms, the information extracted from SNR of noise measurements is limited.
Therefore, the network relies more on the range-Doppler map, which contains rich signal structure information. 
Additionally, due to the randomness in the distribution of false alarms, the contribution of the differences in Doppler channels decreases, while the spatio-temporal coupling becomes critical for the network decision-making. 
Specifically, if the deviation in the coupling between measured Doppler and position information is small, the association is more likely to stem from a target.
Conversely, it is more likely to originate from false alarms.

Thus, through extensive data training, the proposed MFLPN can automatically adjust its association criteria based on data features under different conditions, thereby adaptively selecting corresponding features to make association decisions for observations from various sources. 
Compared to traditional model-driven algorithms, the proposed algorithm effectively integrates multi-dimensional information on different attributes to achieve unified track search and track detection in multiple frames.

\section{SUMMARY}\label{E}
In this work, we proposed the GLP-MFD algorithm to cope with the weak target detection problem.
By constructing observation association graphs with the maximum velocity constraint, the proposed algorithm ensured the successful target associations to mitigate the performance loss due to association failure. 
The MFLPN was designed based on graph neural networks to perform the link prediction task in graphs.
The network fully integrated multi-dimensional information such as Doppler, signal structures, and spatio-temporal coupling, leading to accurate association decisions.
In addition, the proposed algorithm completed track search and track detection in a unified manner, simplifying the processing logic while directly outputting target tracks.
Numerical results indicated that the proposed algorithm enhanced target detection performance while suppressing false tracks, outperforming traditional single-frame and multi-frame detection methods. 
Moreover, the interpretability analysis revealed that the designed network effectively integrated the input features, enabling precise differentiation between target associations and false associations.

To further validate the robustness and versatility of the proposed algorithm, future work will explore its application in more complex environments, including those with interference, clutter, and varying noise levels. 
Additionally, although the computational efficiency of the designed network can be improved through GPU parallel processing, in practical scenarios involving a very large number of detection cells, the graph scale will become substantial.
In such cases, how to reduce the computational complexity of the proposed algorithm requires further exploration.

\begin{IEEEbiography}
[{\includegraphics[width=1in,height=1.25in,clip,keepaspectratio]{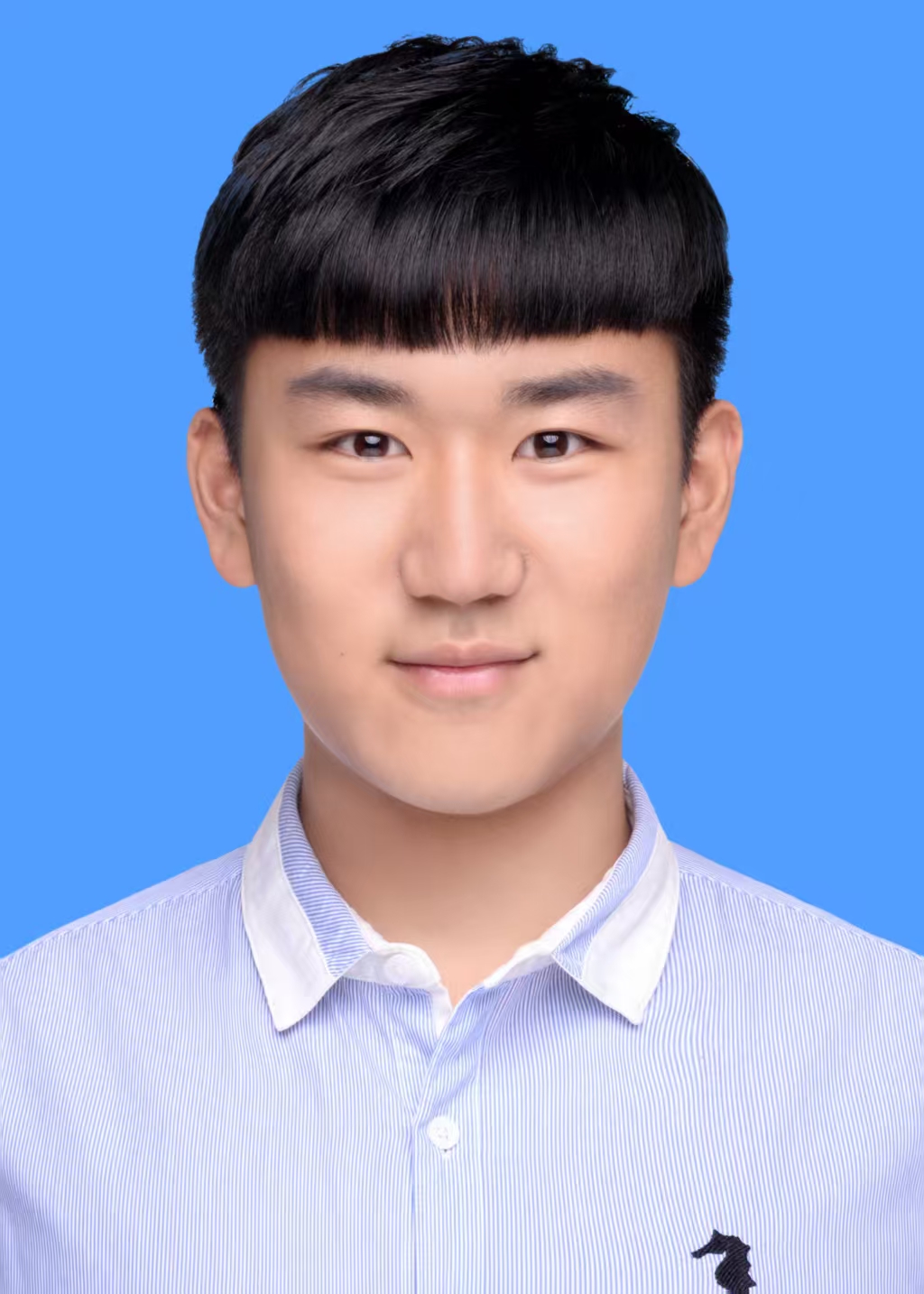}}]
{Zhihao Lin} was born in Shandong, China, in 2000. He received the B.S. degree in electronic information engineering in 2022, from Xidian University, Xi’an, China, where he is currently working toward the Ph.D. degree with the National Key Laboratory of Radar Signal Processing. His research interests include radar signal processing, with a focus on the application of artificial intelligence techniques in radar target detection.
\end{IEEEbiography}

\begin{IEEEbiography}
[{\includegraphics[width=1in,height=1.25in,clip,keepaspectratio]{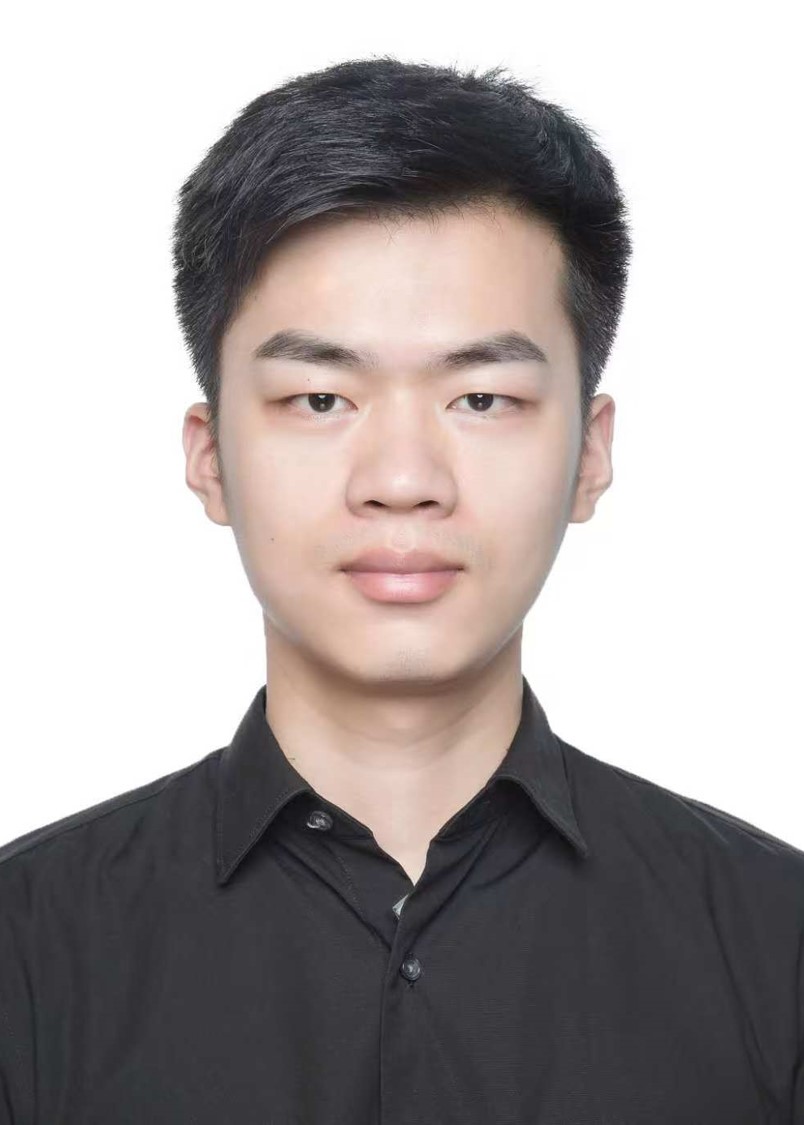}}]
{Chang Gao} (Member, IEEE) received the B.S. and Ph.D. degrees in electrical engineering from Xidian University, Xi’an, China, in 2015 and 2021, respectively. From 2022 to 2024, he was a Postdoctoral Fellow with the Department of Computer Science, City University of Hong Kong. Currently, he is an Associate Professor with the National Key Laboratory of Radar Signal Processing, Xidian University. His research interests include radar target detection and tracking, with a focus on applying optimization and artificial intelligence techniques.
\end{IEEEbiography}

\begin{IEEEbiography}[{\includegraphics[width=1in,height=1.25in,clip,keepaspectratio]{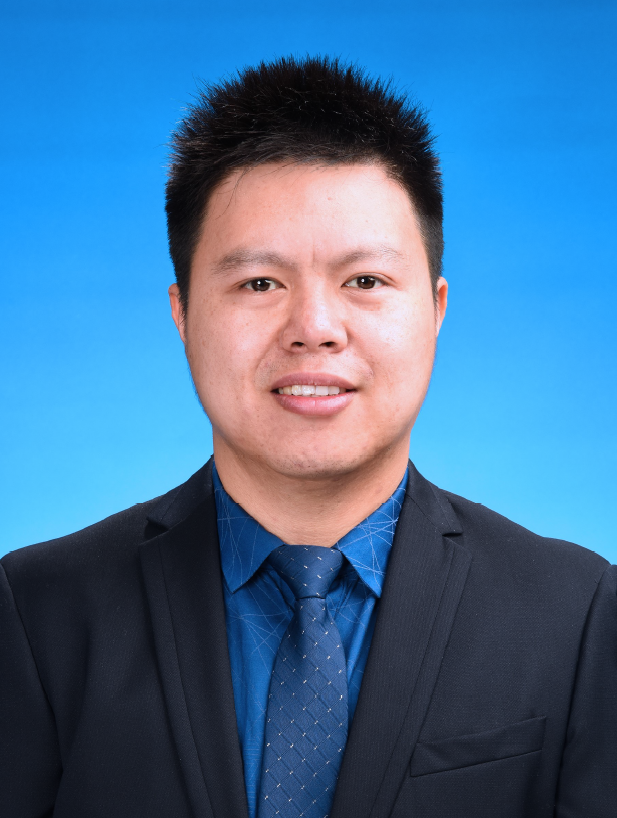}}] 
{Junkun Yan} (Senior Member, IEEE) was born in Sichuan, China, 1987. He received the B.S. and Ph.D. degrees in electronics engineering from Xidian University, Xi’an, China, in 2009 and 2015, respectively. He is currently a Professor with the National Key Laboratory of Radar Signal Processing, Xidian University. He has authored and co-authored more than 100 scientific articles in refereed journals, including IEEE Transactions on Signal Processing, Information Fusion, Signal Processing, and IEEE Transactions on Aerospace and Electronic Systems. His research interests include adaptive signal processing, target tracking, and radar resource allocation. He was a recipient of the Excellent Doctoral thesis Award from the SHANNXI Institute of Electronics in 2017, the Young Talent fund of China Association for Science and Technology in 2020, and the National Youth Talent Support Program in 2023. He is currently an Associate Editor for IEEE Transactions on Aerospace and Electronic Systems and Signal Processing.
\end{IEEEbiography}

\begin{IEEEbiography}
[{\includegraphics[width=1in,height=1.25in,clip,keepaspectratio]{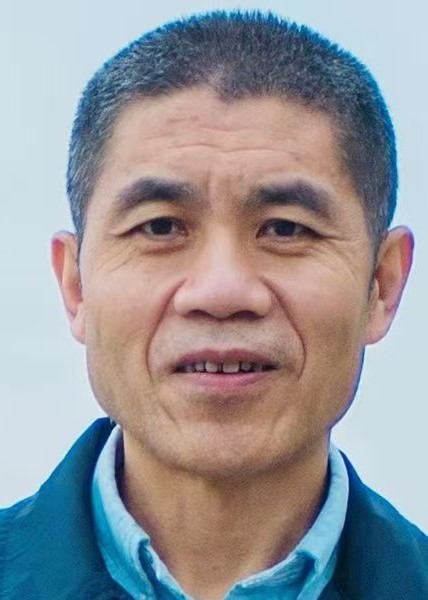}}]
{Qingfu Zhang} (Fellow, IEEE) received the B.Sc. degree in mathematics from Shanxi University, China, in 1984, the M.Sc. degree in applied mathematics and the Ph.D. degree in information engineering from Xidian University, China, in 1991 and 1994, respectively. He is a Chair Professor of computational intelligence with the Department of Computer Science, City University of Hong Kong. He heads a research group with a focus on metaheuristics and artificial intelligence. His MOEA/D algorithms have been widely studied and used in many application fields. He is a Web of Science Highly Cited Researcher in computer science for eight times since 2016.
\end{IEEEbiography}

\begin{IEEEbiography}
[{\includegraphics[width=1in,height=1.25in,clip,keepaspectratio]{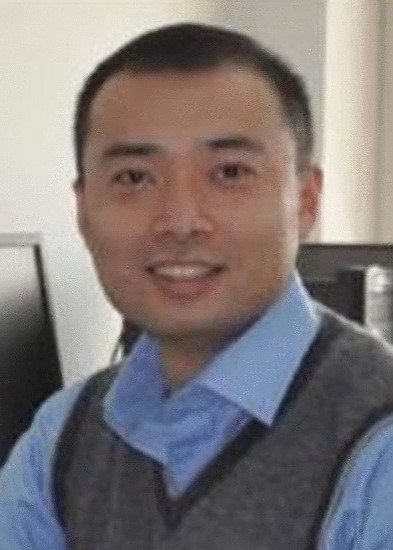}}]
{Bo Chen} (Senior Member, IEEE) received the B.S., M.S., and Ph.D. degrees in electronic engineering from Xidian University, Xi’an, China, in 2003, 2006, and 2008, respectively. From 2008 to 2012, he was a Postdoctoral Fellow, a Research Scientist, and a Senior Research Scientist with the Department of Electrical and Computer Engineering, Duke University, Durham, NC, USA. Since 2013, he has been a Professor with the National Laboratory for Radar Signal Processing, Xidian University. His current research interests include statistical machine learning, statistical signal processing, and radar automatic target detection and recognition. Dr. Chen received the Honorable Mention for the 2010 National Excellent Doctoral Dissertation Award.
\end{IEEEbiography}

\begin{IEEEbiography}
[{\includegraphics[width=1in,height=1.25in,clip,keepaspectratio]{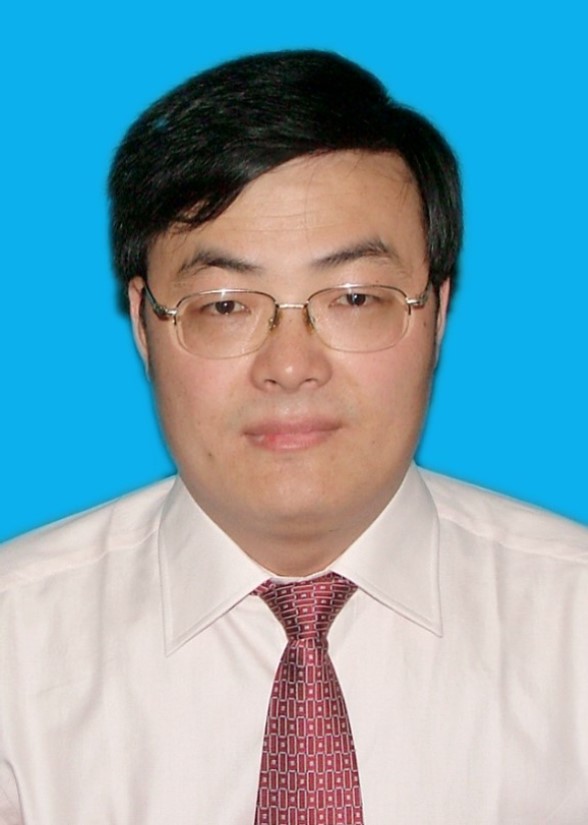}}]
{Hongwei Liu} (Senior Member, IEEE) received the M.S. and Ph.D. degrees in electronic engineering from Xidian University, Xi’an, China, in 1995 and 1999, respectively. He worked with the National Key Laboratory of Radar Signal Processing, Xidian University. From 2001 to 2002, he was a Visiting Scholar with the Department of Electrical and Computer Engineering, Duke University, Durham, NC, USA. Currently, he is a Professor with the National Key Laboratory of Radar Signal Processing and the Vice Principal of Xidian University. His research interests include radar automatic target recognition, radar signal processing, and adaptive signal processing.
\end{IEEEbiography}


\begin{thebibliography}{[34]}
\setcounter{enumiv}{0}

\bibitem{ref1}M.A. Richards
\newblock \emph{Fundamentals of Radar Signal Processing}, 2nd ed. New York, NY, USA: McGraw-Hill, 2014, pp. 406--409.

\bibitem{ref2}\emph{Radar Handbook}, 3rd ed.
\newblock McGraw-Hill, New York, NY, USA, 2008.

\bibitem{ref3}R. Guo, Y. Yuan, and T. Quan
\newblock Adaptive modified hough transform track initiator for HFSWR tracking of fast and small targets
\newblock \emph{J. Syst. Eng. Electron.}, vol. 16, no. 2, pp. 316--320, Jun. 2005.

\bibitem{ref4}A. Farina and F. A. Studer
\newblock \emph{Radar data processing}. Hoboken, NJ, USA: Wiley, 1985.

\bibitem{ref5}P. Wei, J. Zeidler and W. Ku
\newblock Analysis of multiframe target detection using pixel statistics
\newblock \emph{IEEE Trans. Aerosp. Electron. Syst.}, vol. 31, no. 1, pp. 238--247, Jan. 1995, doi: \href{https://dx.doi.org/10.1109/7.366306}{10.1109/7.366306}.


\bibitem{ref7}S. Buzzi, M. Lops and L. Venturino
\newblock Track-before-detect procedures for early detection of moving target from airborne radars
\newblock \emph{IEEE Trans. Aerosp. Electron. Syst.}, vol. 41, no. 3, pp. 937--954, Jul. 2005, doi: \href{https://dx.doi.org/10.1109/TAES.2005.1541440}{10.1109/TAES.2005.1541440}.

\bibitem{ref_Orlando_1}D. Orlando, G. Ricci and Y. Bar-Shalom
\newblock Track-Before-Detect Algorithms for Targets with Kinematic Constraints
\newblock \emph{IEEE Trans. Aerosp. Electron. Syst.}, vol. 47, no. 3, pp. 1837--1849, Jul. 2011, doi: \href{https://dx.doi.org/10.1109/TAES.2011.5937268}{10.1109/TAES.2011.5937268}.

\bibitem{ref_Orlando_2}D. Orlando, L. Venturino, M. Lops and G. Ricci
\newblock Track-Before-Detect Strategies for STAP Radars
\newblock \emph{IEEE Trans. Signal Process.}, vol. 58, no. 2, pp. 933--938, Feb. 2010, doi: \href{https://dx.doi.org/10.1109/TSP.2009.2032991}{10.1109/TSP.2009.2032991}.

\bibitem{ref8}H. Im and T. Kim
\newblock Optimization of multiframe target detection schemes
\newblock \emph{IEEE Trans. Aerosp. Electron. Syst.}, vol. 35, no. 1, pp. 176--187, Jan. 1999, doi: \href{https://dx.doi.org/10.1109/7.745690}{10.1109/7.745690}.


\bibitem{ref10}Y. Barniv
\newblock Dynamic Programming Solution for Detecting Dim Moving Targets
\newblock \emph{IEEE Trans. Aerosp. Electron. Syst.}, vol. AES-21, no. 1, pp. 144--156, Jan. 1985, doi: \href{https://dx.doi.org/10.1109/TAES.1985.310548}{10.1109/TAES.1985.310548}.

\bibitem{ref11}J. Arnold, S. W. Shaw and H. Pasternack
\newblock Efficient target tracking using dynamic programming
\newblock \emph{IEEE Trans. Aerosp. Electron. Syst.}, vol. 29, no. 1, pp. 44--56, Jan. 1993, doi: \href{https://dx.doi.org/10.1109/7.249112}{10.1109/7.249112}.

\bibitem{ref12}W. Li, W. Yi, M. Wen and D. Orlando 
\newblock Multi-PRF and multi-frame track-before-detect algorithm in multiple PRF radar system
\newblock \emph{Signal Process.}, vol. 174, p. 107648, Sep. 2020, doi:
\href{https://dx.doi.org/10.1016/j.sigpro.2020.107648}{10.1016/j.sigpro.2020.107648}.

\bibitem{ref13}W. Li, W. Yi, L. Kong and K. C. Teh
\newblock An Efficient Track-Before-Detect for Multi-PRF Radars With Range and Doppler Ambiguities
\newblock \emph{IEEE Trans. Aerosp. Electron. Syst.}, vol. 58, no. 5, pp. 4083--4100, Oct. 2022, doi:
\href{https://dx.doi.org/10.1109/TAES.2022.3158633}{10.1109/TAES.2022.3158633}.


\bibitem{ref15}P. Uruski and M. Sankowski
\newblock On estimation of performance of track-before-detect algorithm for 3D stacked-beam radar
\newblock In \emph{Proc. Int. Conf. Microw., Radar, Wireless Commun.}, Warsaw, Poland, 2004, pp. 97--100.

\bibitem{ref16}B. D. Carlson, E. D. Evans and S. L. Wilson
\newblock Search radar detection and track with the Hough transform. Part I: System Concept
\newblock \emph{IEEE Trans. Aerosp. Electron. Syst.}, vol. 30, no. 1, pp. 102--108, Jan. 1994, doi: \href{https://dx.doi.org/10.1109/7.250410}{10.1109/7.250410}.

\bibitem{ref17}D. Orlando, F. Ehlers and G. Ricci
\newblock Track-before-detect algorithms for bistatic sonars
\newblock In \emph{Proc. Int. Workshop Cognitive Inf. Process.}, Elba, Italy, 2010, pp. 180--185.

\bibitem{ref18}E. Grossi, L. Venturino and M. Lops
\newblock A two-step multi-frame detection procedure for radar systems
\newblock In \emph{Proc. Int. Conf. Inf. Fusion}, Singapore, 2012, pp. 1196--1201.

\bibitem{ref19}E. Grossi, M. Lops and L. Venturino
\newblock A Novel Dynamic Programming Algorithm for Track-Before-Detect in Radar Systems
\newblock \emph{IEEE Trans. Signal Process.}, vol. 61, no. 10, pp. 2608--2619, May 2013, doi: \href{https://dx.doi.org/10.1109/TSP.2013.2251338}{10.1109/TSP.2013.2251338}.

\bibitem{ref20}E. Grossi, M. Lops and L. Venturino
\newblock Track-before-detect for multiframe detection with censored observations
\newblock \emph{IEEE Trans. Aerosp. Electron. Syst.}, vol. 50, no. 3, pp. 2032--2046, Jul. 2014, doi: \href{https://dx.doi.org/10.1109/TAES.2013.130148}{10.1109/TAES.2013.130148}.

\bibitem{ref21}S.-W. Yeom, T. Kirubarajan and Y. Bar-Shalom
\newblock Track segment association, fine-step IMM and initialization with Doppler for improved track performance
\newblock \emph{IEEE Trans. Aerosp. Electron. Syst.}, vol. 40, no. 1, pp. 293--309, Jan. 2004, doi: \href{https://dx.doi.org/10.1109/TAES.2004.1292161}{10.1109/TAES.2004.1292161}.

\bibitem{ref22}F. Kural, F. Ankan, O. Arikan and M. Efe
\newblock Incorporating Doppler Velocity Measurement for Track Initiation and Maintenance
\newblock In \emph{Proc. IEE Seminar Target Tracking Algorithms Appl.}, Birmingham, AL, USA, 2006, pp. 107--114.

\bibitem{ref23}D. Mušicki, T. L. Song, H. H. Lee and D. Nešić
\newblock Correlated Doppler-assisted target tracking in clutter
\newblock \emph{IET Radar, Sonar Navig.}, vol. 7, no. 1, pp. 94--100, Jan. 2013, doi: 
\href{https://dx.doi.org/10.1049/iet-rsn.2012.0115}{10.1049/iet-rsn.2012.0115}.

\bibitem{ref24}C. Gao, J. Yan, X. Peng, B. Chen and H. Liu
\newblock Intelligent multiframe detection aided by Doppler information and a deep neural network
\newblock \emph{Inf. Sci.}, vol. 593, pp. 432--448, May 2022, doi: 
\href{https://dx.doi.org/10.1016/j.ins.2022.01.029}{10.1016/j.ins.2022.01.029}.

\bibitem{ref25}Y. Bar-Shalom
\newblock \emph{Tracking and data association}. San Diego, CA, USA: Academic Press Professional, Inc., 1987.

\bibitem{ref26}W. Xiong, Y. Lu, J. Song and X. Chen
\newblock A Two-Stage Track-before-Detect Method for Non-Cooperative Bistatic Radar Based on Deep Learning
\newblock \emph{Remote Sens.}, vol. 15, no. 15, p. 3757, Jul. 2023, doi: 
\href{https://dx.doi.org/10.3390/rs15153757}{10.3390/rs15153757}.

\bibitem{ref27}Y. Chen, Y. Wang, F. Qu and W. Li
\newblock A Graph-Based Track-Before-Detect Algorithm for Automotive Radar Target Detection
\newblock \emph{IEEE Sens. J.}, vol. 21, no. 5, pp. 6587--6599, Mar. 2021, doi: 
\href{https://dx.doi.org/10.1109/JSEN.2020.3042079}{10.1109/JSEN.2020.3042079}.

\bibitem{ref28}F. Scarselli, M. Gori, A. C. Tsoi, M. Hagenbuchner and G. Monfardini
\newblock The Graph Neural Network Model
\newblock \emph{IEEE Trans. Neural Netw.}, vol. 20, no. 1, pp. 61--80, Jan. 2009, doi: 
\href{https://dx.doi.org/10.1109/TNN.2008.2005605}{10.1109/TNN.2008.2005605}.

\bibitem{ref29}L. Wu, Y. Chen, K. Shen, X. Guo, H. Gao, S. Li \emph{et al.}
\newblock \emph{Graph Neural Networks for Natural Language Processing: A Survey}. Norwell, MA, USA: Now Publishers Inc., 2023.

\bibitem{ref30}Z. Wu, S. Pan, F. Chen, G. Long, C. Zhang and P. S. Yu
\newblock A Comprehensive Survey on Graph Neural Networks
\newblock \emph{IEEE Trans. Neural Netw. Learn. Syst.}, vol. 32, no. 1, pp. 4--24, Jan. 2021, doi: 
\href{https://dx.doi.org/10.1109/TNNLS.2020.2978386}{10.1109/TNNLS.2020.2978386}.


\bibitem{ref32}A. Scotti, N. N. Moghadam, D. Liu, K. Gafvert and J. Huang
\newblock Graph Neural Networks for Massive MIMO Detection
\newblock 2020, \emph{arXiv:2007.05703v1}.

\bibitem{ref33}F. Fent, P. Bauerschmidt and M. Lienkamp
\newblock RadarGNN: Transformation Invariant Graph Neural Network for Radar-based Perception
\newblock In \emph{Proc. IEEE Conf. Comput. Vis. Pattern Recognit. Workshops}, Vancouver, BC, Canada, 2023, pp. 182--191.

\bibitem{ref34}N. Su, X. Chen, J. Guan, Y. Huang, X. Wang and Y. Xue
\newblock Radar Maritime Target Detection via Spatial–Temporal Feature Attention Graph Convolutional Network
\newblock \emph{IEEE Trans. Geosci. Remote Sens.}, vol. 62, no. 5102615, pp. 1--15, Jan. 2024, doi: 
\href{https://dx.doi.org/10.1109/TGRS.2024.3358862}{10.1109/TGRS.2024.3358862}.

\bibitem{ref35}C. Gao, J. Yan, B. Chen, P. K. Varshney, T. Jia and H. Liu
\newblock Data association for maneuvering targets through a combined siamese network and XGBoost model
\newblock \emph{Signal Process.}, vol. 211, no. 109086, pp. 1--12, Oct. 2023, doi: 
\href{https://dx.doi.org/10.1016/j.sigpro.2023.109086}{10.1016/j.sigpro.2023.109086}.

\bibitem{ref36}B. P. Chamberlain, S. Shirobokov, E. Rossi, F. Frasca, T. Markovich, N. Y. Hammerla \emph{et al.}
\newblock Graph Neural Networks for Link Prediction with Subgraph Sketching
\newblock In \emph{Int. Conf. Learn. Representations}, Kigali, Rwanda, 2023, pp. 1--27.

\bibitem{ref37}M. Gao, P. Jiao, R. Lu, H. Wu, Y. Wang and Z. Zhao
\newblock Inductive Link Prediction via Interactive Learning Across Relations in Multiplex Networks
\newblock \emph{IEEE Trans. Comput. Social Syst.}, vol. 11, no. 3, pp. 3118--3130, Jun. 2024, doi: 
\href{https://dx.doi.org/10.1109/TCSS.2022.3176928}{10.1109/TCSS.2022.3176928}.

\bibitem{ref38}W. Liu, W. Xie, J. Liu and Y. Wang
\newblock Adaptive Double Subspace Signal Detection in Gaussian Background—Part I: Homogeneous Environments
\newblock \emph{IEEE Trans. Signal Process.}, vol. 62, no. 9, pp. 2345--2357, May 2014, doi:
\href{https://dx.doi.org/10.1109/TSP.2014.2309556}{10.1109/TSP.2014.2309556}.

\bibitem{ref39}W. Liu, J. Liu, T. Liu, H. Chen and Y. -L. Wang
\newblock Detector Design and Performance Analysis for Target Detection in Subspace Interference
\newblock \emph{IEEE Signal Process. Lett.}, vol. 30, pp. 618--622, Apr. 2023, doi:
\href{https://dx.doi.org/10.1109/LSP.2023.3270080}{10.1109/LSP.2023.3270080}.


\bibitem{refdet2}J. Z. Hare, Y. Liang, L. M. Kaplan and V. V. Veeravalli
\newblock Bayesian Two-Sample Hypothesis Testing Using the Uncertain Likelihood Ratio: Improving the Generalized Likelihood Ratio Test
\newblock \emph{IEEE Trans. Signal Process.}, vol. 73, pp. 1410--1425, Mar. 2025, doi:
\href{https://dx.doi.org/10.1109/TSP.2025.3546169}{10.1109/TSP.2025.3546169}.


\bibitem{refdet3}T. Wang, C. Yin, D. Xu, C. Hao, D. Orlando and G. Ricci
\newblock Analysis of MIMO Radar Detection Algorithms With Location Capabilities: CFAR Property and Selectivity
\newblock \emph{IEEE Trans. Aerosp. Electron. Syst.}, vol. 61, no. 2, pp. 5426--5435, Apr. 2025, doi: \href{https://dx.doi.org/10.1109/TAES.2024.3488688}{10.1109/TAES.2024.3488688}.


\bibitem{refdet4}Z. Xu, W. Liu, C. Wu, Q. Du and J. Liu
\newblock Statistical Performance of Generalized Direction Detectors with Known Spatial Steering Vector
\newblock 2025, \emph{arXiv:2505.03076v2}.



\bibitem{ref40}D. Gaifulina and I. Kotenko
\newblock Selection of Deep Neural Network Models for IoT Anomaly Detection Experiments
\newblock In \emph{Proc. Euromicro Int. Conf. Parallel, Distrib. Network-Based Process.}, Valladolid, Spain, 2021, pp. 260--265.


\bibitem{refdet1}D. Xiao, W. Liu, J. Liu, Y. Wu, Q. Du and X. Hua
\newblock Bayesian Rao test for distributed target detection in interference and noise with limited training data
\newblock \emph{Sci. China Inf. Sci.}, May 2025, doi:
\href{https://www.sciengine.com/SCIS/doi/10.1007/s11432-024-4422-3}{10.1007/s11432-024-4422-3}.


\bibitem{ref41}C. Gao, J. Yan, X. Peng and H. Liu
\newblock Signal structure information-based target detection with a fully convolutional network
\newblock \emph{Inf. Sci.}, vol. 576, pp. 345--354, Oct. 2021, doi: 
\href{https://dx.doi.org/10.1016/j.ins.2021.06.066}{10.1016/j.ins.2021.06.066}.


\bibitem{ref42}W. Liu, Y. Wu, Q. Jiang, J. Liu, S. Xu and P. Gong
\newblock Eigenvalue-based distributed target detection in compound-Gaussian clutter
\newblock \emph{Sci. China Inf. Sci.}, Feb. 2025, doi:
\href{https://www.sciengine.com/SCIS/doi/10.1007/s11432-024-4319-5}{10.1007/s11432-024-4319-5}.


\bibitem{ref43}J. Gilmer, S. S. Schoenholz, P. F. Riley, O. Vinyals and G. E. Dahl
\newblock Neural message passing for Quantum chemistry
\newblock In \emph{Pro. Int. Con. Mach. Learn.}, Sydney, NSW, Australia, 2017, pp. 1263-–1272.

\bibitem{ref44}P. Veličković, G. Cucurull, A. Casanova, A. Romero, P. Liò and Y. Bengio
\newblock Graph Attention Networks 
\newblock In \emph{Int. Conf. Learn. Representations}, Vancouver, Canada, 2018, pp. 1--12.

\bibitem{ref45}S.Brody, U. Alon and E. Yahav
\newblock How Attentive are Graph Attention Networks?
\newblock In \emph{Int. Conf. Learn. Representations}, Virtual, 2022, pp. 1--26.

\bibitem{ref_add_1}D. Yu, Y. Zhou, S. Zhang, W. Li, M. Small and K. Shang
\newblock Information cascade prediction of complex networks based on physics-informed graph convolutional network
\newblock \emph{New J. Phys.}, vol. 26, no. 1, p. 013031, Jan. 2024, 
doi: \href{https://dx.doi.org/10.1088/1367-2630/ad1b29}{10.1088/1367-2630/ad1b29}.

\bibitem{ref46}P. H. C. Avelar, A. R. Tavares, T. L. T. da Silveira, C. R. Jung and L. C. Lamb
\newblock Superpixel Image Classification with Graph Attention Networks
\newblock In \emph{Proc. SIBGRAPI Conf. Graph., Patterns Imag.}, Porto de Galinhas, Brazil, 2020, pp. 203--209.

\bibitem{ref47}Q. Li, Y. Shang, X. Qiao and W. Dai
\newblock Heterogeneous Dynamic Graph Attention Network
\newblock In \emph{Proc. IEEE Int. Conf. Knowl. Graph}, Nanjing, China, 2020, pp. 404--411.

\bibitem{ref48}I. Goodfellow, Y. Bengio and A. Courville
\newblock \emph{Deep Learning}. Cambridge, MA, USA: MIT Press, 2016.

\bibitem{ref_add_2}A. Paszke, S. Gross, F. Massa, A. Lerer, J. Bradbury, G. Chanan \emph{et al.}
\newblock PyTorch: An Imperative Style, High-Performance Deep Learning Library
\newblock In \emph{NeurIPS}, Vancouver Canada, 2019, pp. 8024--8035.

\bibitem{ref_add_3}D. P. Kingma and J. Ba
\newblock Adam: A Method for Stochastic Optimization
\newblock In \emph{Int. Conf. Learn. Representations}, San Diego, CA, USA, 2015, pp. 13--27.

\bibitem{ref_add_4}T. N. Kipf and M. Welling
\newblock Semi-Supervised Classification with Graph Convolutional Networks
\newblock In \emph{Int. Conf. Learn. Representations}, Toulon, France, 2017, pp. 1--14.


\bibitem{ref_hyper}S. Roy, R. Mehera, R. K. Pal and S. K. Bandyopadhyay
\newblock Hyperparameter optimization for deep neural network models: a comprehensive study on methods and techniques
\newblock \emph{Innovations Syst. Softw. Eng.}, Oct. 2023, 
doi: \href{https://dx.doi.org/10.1007/s11334-023-00540-3}{10.1007/s11334-023-00540-3}.


\bibitem{ref49}B. Ristic, B. -N. Vo, D. Clark and B. -T. Vo
\newblock A Metric for Performance Evaluation of Multi-Target Tracking Algorithms
\newblock \emph{IEEE Trans. Signal Process.}, vol. 59, no. 7, pp. 3452--3457, Jul. 2011, doi: \href{https://dx.doi.org/10.1109/TSP.2011.2140111}{10.1109/TSP.2011.2140111}.

\bibitem{ref50}D. Schuhmacher, B. -T. Vo and B. -N. Vo
\newblock A Consistent Metric for Performance Evaluation of Multi-Object Filters
\newblock \emph{IEEE Trans. Signal Process.}, vol. 56, no. 8, pp. 3447--3457, Aug. 2008, doi: \href{https://dx.doi.org/10.1109/TSP.2008.920469}{10.1109/TSP.2008.920469}.



\bibitem{ref_distr}C. Gao, Q. Zhang, P. K. Varshney, X. Lin and H. Liu
\newblock A Distributed Multi-Objective Detection Method for Multi-Sensor Systems With Unknown Local SNR
\newblock \emph{IEEE Trans. Signal Process.}, vol. 73, pp. 649--663, Jan. 2025, doi: \href{https://dx.doi.org/10.1109/TSP.2025.3533275}{10.1109/TSP.2025.3533275}.


\bibitem{ref51}M. Ancona, E. Ceolini, A. C. Öztireli and M. Gross
\newblock A unified view of gradient-based attribution methods for Deep Neural Networks
\newblock In \emph{NIPS Workshop Interpreting, Explaining, Visualizing Deep Learn.}, Long Beach, CA, USA, 2017, pp. 1--11.

\bibitem{ref52}L. Breiman 
\newblock Random Forests
\newblock \emph{Mach. Learn.}, vol. 45, pp. 5--32, Oct. 2001, doi: 
\href{https://dx.doi.org/10.1023/A:1010933404324}{10.1023/A:1010933404324}.

\bibitem{ref53}F. K. Ewald, L. Bothmann, M. N. Wright, B. Bischl, G. Casalicchio and G. König
\newblock A Guide to Feature Importance Methods for Scientific Inference
\newblock 2024, \emph{arXiv:2404.12862}.

\bibitem{ref54}N .J. Cox 
\newblock Speaking Stata: Creating and Varying Box Plots
\newblock \emph{Stata J.}, vol. 9, no. 3, pp. 478--496, Sep. 2009, doi: 
\href{https://dx.doi.org/10.1177/1536867X0900900309}{10.1177/1536867X0900900309}.


\end{thebibliography}
\end{document}